\documentclass[fleqn,usenatbib]{mnras}

\usepackage{newtxtext,newtxmath}
\usepackage[T1]{fontenc}

\DeclareRobustCommand{\VAN}[3]{#2}
\let\VANthebibliography\thebibliography
\def\thebibliography{\DeclareRobustCommand{\VAN}[3]{##3}\VANthebibliography}

\usepackage{graphicx}	
\usepackage{amsmath}	
\usepackage{xspace}
\usepackage{tabularx}
\usepackage{booktabs}

\usepackage{etoolbox}
\makeatletter
\patchcmd{\NAT@citex}
  {\@citea\NAT@hyper@{\NAT@nmfmt{\NAT@nm}\NAT@date}}
  {\@citea\NAT@nmfmt{\NAT@nm}\NAT@hyper@{\NAT@date}}
  {}
  {}
\patchcmd{\NAT@citex}
  {\@citea\NAT@hyper@{%
     \NAT@nmfmt{\NAT@nm}%
     \hyper@natlinkbreak{\NAT@aysep\NAT@spacechar}{\@citeb\@extra@b@citeb}%
     \NAT@date}}
  {\@citea\NAT@nmfmt{\NAT@nm}%
   \NAT@aysep\NAT@spacechar%
   \NAT@hyper@{\NAT@date}}
  {}
  {}
\patchcmd{\NAT@citex}
  {\@citea\NAT@hyper@{%
     \NAT@nmfmt{\NAT@nm}%
     \hyper@natlinkbreak{\NAT@spacechar\NAT@@open\if*#1*\else#1\NAT@spacechar\fi}%
       {\@citeb\@extra@b@citeb}%
     \NAT@date}}
  {\@citea\NAT@nmfmt{\NAT@nm}%
   \NAT@spacechar\NAT@@open\if*#1*\else#1\NAT@spacechar\fi%
   \NAT@hyper@{\NAT@date}}
  {}
  {}
\makeatother

\newcommand{\req}[0]{\ensuremath{R_{\mathrm{eq}}}\xspace}
\newcommand{\sat}[0]{\ensuremath{P_{s \times s}}\xspace}
\newcommand{\cen}[0]{\ensuremath{P_{c \times c}}\xspace}
\newcommand{\censat}[0]{\ensuremath{P_{c \times s}}\xspace}
\newcommand{\hi}{${\rm H}$\,{\sc i}\xspace}

\newcommand{\gr}[0]{$g\!$ - $\!r$\xspace}
\newcommand{\units}[2]{\textrm{#1}$^{\mathrm{#2}}$}
\newcommand{\srm}[1]{_{\rm{#1}}}
\newcommand{\mpch}[0]{\textit{h}$^{-1}$ cMpc\xspace}
\newcommand{\hmpc}[0]{\textit{h} cMpc$^{-1}$\xspace}
\newcommand{\hib}{\hi $\times$ Blue\xspace}
\newcommand{\hir}{\hi $\times$ Red\xspace}
\newcommand{\hia}{\hi $\times$ Galaxy\xspace}
\newcommand{\mhi}{\ensuremath{M_{\mathrm{HI}}}\xspace}

\defcitealias{DiemerHI2018}{D18}
\defcitealias{V-NIngred21cm2018}{VN18}

\title[HI and Galaxy Colour]{Atomic Hydrogen Shows its True Colours: Correlations between HI and Galaxy Colour in Simulations}

\author[Osinga et al.]{
Calvin K. Osinga$^{1}$\thanks{E-mail: cosinga@umd.edu},
Benedikt Diemer$^{1}$,
Francisco Villaescusa-Navarro $^{2}$, \newauthor
Elena D'Onghia $^{3, 4}$,
Peter Timbie$^{4}$
\\
$^{1}$Department of Astronomy, University of Maryland - College Park, 4296 Stadium Dr, 20742, College Park, MD, USA\\
$^{2}$Center for Computational Astrophysics, Flatiron Institute, 162 5th Avenue, 10010, New York, NY, USA\\
$^{3}$Department of Astronomy, University of Wisconsin - Madison, 475 North Charter Street, Madison, WI 53706, USA\\
$^{4}$Department of Physics, University of Wisconsin - Madison, 1150 University Avenue, Madison, WI 53706, USA\\
}

\pubyear{2023}

\begin{document}

\label{firstpage}
\pagerange{\pageref{firstpage}--\pageref{lastpage}}
\maketitle

\begin{abstract}
Intensity mapping experiments are beginning to measure the spatial distribution of neutral atomic hydrogen HI to constrain cosmological parameters and the large-scale distribution of matter. However, models of the behaviour of HI as a tracer of matter are complicated by galaxy evolution. In this work, we examine the clustering of HI in relation to galaxy colour, stellar mass, and HI mass in IllustrisTNG at $z$ = 0, 0.5, and 1. We compare the HI-red and HI-blue galaxy cross-power spectra, finding that HI-red has an amplitude 1.5 times greater than HI-blue at large scales. The cross-power spectra intersect at $\approx 3$ Mpc in real space and $\approx 10$ Mpc in redshift space, consistent with $z \approx 0$ observations. We show that HI clustering increases with galaxy HI mass and depends weakly on detection limits in the range $\mhi \leq 10^8 M_\odot$. In terms of $M_\star$, we find massive blue galaxies cluster more than less massive ones. Massive red galaxies, however, cluster the weakest amongst red galaxies. These opposing trends arise from central-satellite compositions. Despite these $M_\star$ trends, we find that the cross-power spectra are largely insensitive to detection limits in galaxy surveys. Counter-intuitively, all auto and cross-power spectra for red and blue galaxies and HI decrease with time at all scales. We demonstrate that processes associated with quenching contribute to this trend. The complex interplay between HI and galaxies underscores the importance of understanding baryonic effects when interpreting the large-scale clustering of HI, blue, and red galaxies at $z \leq 1$.
\end{abstract}




\section{Introduction}

In the current paradigm of structure formation, gravity transforms small primordial density fluctuations in our nearly homogeneous Universe into the cosmic web. Within this web, overdense regions of dark matter known as haloes emerge via gravitational instabilities. Over time, baryons sink into gravitational potential wells of haloes and form galaxies \citep{WhiteStructure1978}. Consequently, the clustering of galaxies is shaped by the cosmology responsible for the large-scale distribution of haloes \citep{JenkinsCosmoParam1998, EisensteinSDSSBAO2005, Reddick2014} and how galaxies occupy haloes \citep{ZhengHOD2007}. Studies of the clustering in galaxy surveys therefore provide insight into the behaviour of dark matter and dark energy and the influence of dark matter haloes on galaxy properties.
 
Galaxy surveys measure the large-scale distribution of galaxies via their starlight and thus focus on the stellar properties of galaxies. However, the gas properties of a galaxy are also critical in galaxy evolution, as they are tightly linked to a galaxy's star-formation rate \citep[SFR,][]{K-SRel1959, K-SRel1998}. Fortunately, future experiments that map the distribution of neutral atomic hydrogen (\hi) can probe the link between the gas properties of galaxies and their host halo, called the \hi-galaxy-halo connection \citep{Guo2017HImassClustering, LiHIModelColor2022}. Moreover, these maps can also constrain cosmological parameters as a competitive alternative to galaxy surveys \citep{Ansari21cm2018}. By sacrificing angular resolution, 21cm intensity mapping experiments can improve the signal-to-noise ratio and, in principle, observe the structure of the universe quickly and efficiently \citep{LeoHIasprobe2019}. Projects intended for this purpose such as CHIME \citep{CHIME2014}, Tianlai \citep{Tianlai2021}, HIRAX \citep{HIRAX2016}, and SKA \citep{SKA2009} are largely still in their proof-of-concept phase. Just recently, authors in \citet{Paul2023MEERKATHIauto} claim to have successfully detected the \hi signal in auto-correlation. 

The spatial distribution of \hi is shaped by the confluence of matter's large-scale structure and how \hi occupies haloes. Consequently, a large quantity of work has been dedicated to understanding the \hi-galaxy-halo connection via studying galaxy clustering as functions of different properties \citep[e.g.,][]{ZehaviLumColCluster2005, Li2012, AndersonHI-color2018, QinHIHOD2022}. One such result is the tendency for ``red'' galaxies to cluster more strongly than ``blue'' galaxies \citep{ZehaviColGalClus2011, Skibba2015, coil2017primus+}. Colour-dependent clustering arises from the propensity for red galaxies to occupy older and more massive haloes, which also tend to be the most clustered via ``assembly bias'' \citep{GaoAgeClustering2005}. A galaxy's colour reflects its star-formation status; galaxies with low SFRs cannot maintain a substantial population of short-lived blue stars, yielding a redder colour on average \citep{TinsleySFR1980, MadauColorSFR1996}. The cessation of star-formation (called quenching) follows from the depletion of cold molecular gas reservoirs \citep{Jimenez2023VirgoK-S}, and thus is particularly relevant to understanding the spatial distribution of \hi. A galaxy's \hi abundance is correlated with its SFR, so the mechanisms that eventually transform a galaxy from blue to red usually also suppress its \hi content \citep{BigielSFR-HI2008, Wang2020}.

Studying the spatial relationship between \hi and galaxy colour furthers our understanding of the processes responsible for quenching. Past works have measured cross-correlations at nearby redshifts ($0 < z < 1$) between \hi and galaxies separated into blue and red populations \citep{ChangDEEP2-GBT2010, MasuiWiggleZ-GBT2013, BigPapa2013, AndersonHI-color2018, WolzeBOSSGBT2021, CunningtonCC2023, jiang2023crosscorrelation}. They find that the abundance of \hi is suppressed in regions within anywhere from 2 Mpc to 9 Mpc of a red galaxy. Physically interpreting these results as contributions from evolving galaxy populations is a formidable analytical task; simulations offer a way to address this challenge. Moreover, the current generation of simulations now produce low-redshift galaxy populations with realistic colour and \hi properties \citep{NelsonTNG2018, DiemerHI2019, Dave2020}, offering the opportunity to illuminate the relationship between galaxy colour and the spatial distribution of \hi.

In this work, we analyse the \hi auto power spectrum and \hi-galaxy cross-power spectra in real and redshift space using the hydrodynamical simulation IllustrisTNG. For each power spectrum, we characterize their dependence on scale, colour, \hi mass, and stellar mass at $z = 0$, $z = 0.5$, and $z = 1$. We find that the \hi distribution is colour-dependent to surprisingly large scales. In an upcoming paper (Osinga et al. in prep.), we will analyse the implications of the large-scale colour-dependence on the cosmological interpretation of \hi intensity maps at low redshift. In this work, we focus on the insight that these cross-correlations provide on galaxy formation and evolution.

The paper is structured as follows. In Section~\ref{section:methods}, we describe the simulated dataset used in our analysis and various mathematical definitions. In Section~\ref{section:results}, we study the clustering of \hi and galaxy colour cross-power spectra and their relationships with redshift, stellar mass, and \hi mass. We then analyse the ramifications of these results in Section~\ref{section:discussion} before concluding in Section~\ref{section:conclusion}. For brevity, some additional figures are referenced but are not included. These are provided on the author's website\footnote{\url{www.calvinosinga.com}}.

\section{Methods}
\label{section:methods}

\subsection{Simulation data}
\label{subsection:TNG}

We base this work on data from the IllustrisTNG suite of cosmological magneto-hydrodynamics simulations \citep{NelsonTNG2018, PillepichTNG2018, SpringelTNG2018, NaimanTNG2018, MarinacciTNG2018, Nelson19}. The suite provides simulations at three resolutions in three different volumes, with side lengths 35 \mpch, 75 \mpch, and 205 \mpch. The simulations adopt the \citet{Planck2015} cosmology with $\Omega_{\rm{m}} = 0.3089$, $\Omega_{\rm{b}} = 0.0486$, $h = 0.6774$ and $\sigma_{\rm{s}} = 0.8159$. 

The boxes are evolved using the moving-mesh code AREPO \citep{SpringelAREPO2010} that calculates gravity with a tree-PM method and magneto-hydrodynamics with a Gudonov scheme using a Voronoi mesh. IllustrisTNG applies sub-grid models for unresolved processes such as star formation, stellar winds, gas cooling, supernovae, and active galactic nuclei \citep[AGN,][]{VogelGalFormModel2013, Weinberger2018}. The parameters to these models are tuned using a small subset of observations to produce a realistic low-redshift galaxy population \citep{PillepichTNGmodel2018}. A Friends-of-Friends (FoF) algorithm \citep{DavisFOF1985} groups particles into dark matter haloes. Overdense substructure within the haloes called ``subhaloes'' are identified using the \textsc{Subfind} algorithm \citep{SpringelSUBFIND2001}.

We primarily use the highest-resolution 75 \hmpc box, called TNG100, at redshifts $z = 0$, $z = 0.5$, and $z = 1$. The largest box, TNG300, offers power spectra at larger scales, but unresolved galaxies contribute a non-negligible proportion of the cosmic \hi (see \citet{V-NIngred21cm2018} and Appendix~\ref{appendix:convergence}) and the smallest box does not reach the scales of interest for 21cm intensity mapping.

\begin{figure}
    \centering
    \includegraphics[width=\linewidth, trim={0cm 0.25cm 0cm 0cm}, clip]{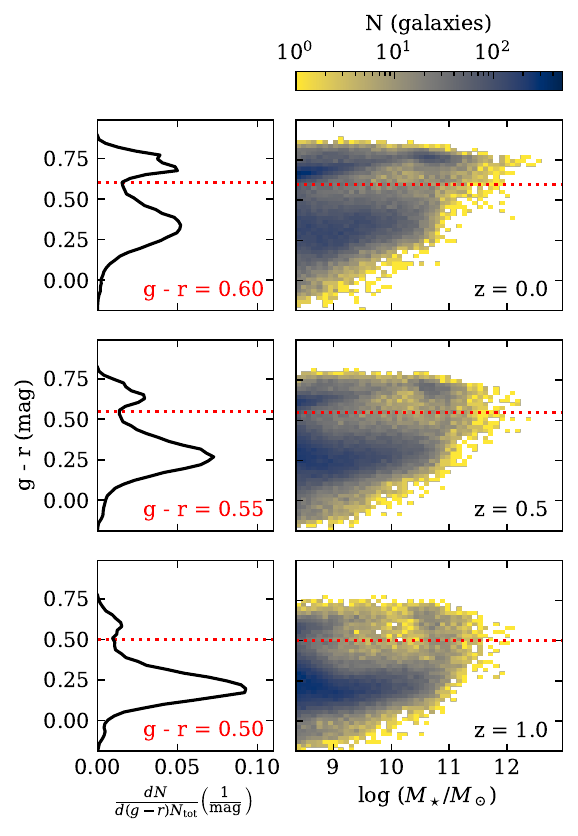}
    \caption{Distribution of rest-frame galaxy colours in TNG100 at $z = 0$ (top), $z = 0.5$ (centre), and $z = 1$ (bottom) without dust attenuation. 2D histograms (right) show the colour-$M_\star$ distribution. 1D colour distributions (left) sum along each \gr bin in the colour-$M_\star$ plane, normalised by the number of galaxies at that redshift, $N\srm{tot}$. The red dotted line represents the bin with the minimum galaxy count between the peaks of the bimodal distribution, chosen to be the threshold that separates galaxies into blue (below line) and red (above line) subpopulations. The galaxy colour distribution in IllustrisTNG lacks dusty, star-forming red galaxies \citep{DonnariUVJTNG2019}. However, the colour distribution at these redshifts is reasonable \citep{Nelson19} since dusty, star-forming galaxies are not thought to form a substantial fraction of red galaxies at $z < 1$ (see text for discussion). The blue cloud is distributed over a wide range of \gr values and reaches a maximum at $M_\star \approx 10^{11} M_\odot$. The red sequence has a tight \gr distribution with larger spread in $M_\star$.}
    \label{fig:gr-stmass}
\end{figure}

We remove galaxies with $M_\star < 2 \times 10^{8}$ $\rm{M}_\odot$ as the data from galaxies with fewer than 100 stellar particles is unreliable. We then separate the galaxies into blue and red using the difference in magnitude in the g- and r-bands, as defined by the Sloan Digital Sky Survey \citep[SDSS, ][]{StoughtonSDSSBands2002}. The resulting colour$-M_\star$ plane is shown in Fig.~\ref{fig:gr-stmass}. For each redshift, we select the \gr value that corresponds to the bin with the minimum count between the two peaks in the colour distribution. Galaxies falling on or above the line are classified as blue and all others as red. We find that the minimum evolves from \gr = 0.60, \gr = 0.55, and \gr = 0.50 at $z = 0$, $z = 0.5$, and $z = 1$, respectively. \citet{NelsonTNG2018} find that IllustrisTNG's \gr distribution roughly matches observations in our redshift range.

However, at high redshift ($z > 1$) IllustrisTNG's colour distribution begins to diverge from observations. IllustrisTNG lacks dusty and star-forming red galaxies \citep{DonnariUVJTNG2019}, an observed phenomenon that is also missing in other simulations such as MUFASA \citep{DaveMUFASAred2017}. Only recently has the follow-up simulation for MUFASA, SIMBA \citep{SIMBA2019}, produced this population \citep{AtkinsSIMBA2022}. We use redshifts $z \leq 1.0$ where the star-forming red population is negligible in IllustrisTNG (online figures), such that nearly all red galaxies are quenched and blue galaxies star-forming. We test the sensitivity of the blue and red galaxy clustering to different \gr thresholds and the effect of dust reddening according to the model from \citet{NelsonTNG2018}. The clustering of both blue and red galaxies is not appreciably impacted on large scales by any reasonably chosen colour threshold or by dust reddening (see Appendix~\ref{appendix:color_cut}). \citet{SpringelTNG2018} compared the clustering of blue and red galaxies from TNG300 to SDSS measurements, finding excellent agreement for blue galaxies at all stellar masses and a slight overestimation of red galaxy clustering at intermediate stellar masses (see their figure 11).

\subsection{Modelling atomic and molecular hydrogen}
\label{subsection:phase}

We adopt the same notation as \citet{DiemerHI2018} to distinguish between the states of hydrogen, involving a consistent set of subscripts for any physical quantity that would be associated with the gas. If $\Sigma$ is the surface density, then $\Sigma_{\rm{gas}}$ is adopted for all gas, $\Sigma_{\rm{H}}$ for all hydrogen, $\Sigma_{\rm{HI+H}_2}$ for neutral hydrogen, $\Sigma_{\rm{HI}}$ for atomic hydrogen, and $\Sigma_{\rm{H}_2}$ for molecular hydrogen. Using this notation, the molecular fraction is defined as
\begin{equation}
    \label{eq:molfrac}
    f_{\rm{mol}} = \frac{M_{\rm{H}_2}}{M_{\rm{HI} + \rm{H}_2}} \,.
\end{equation}
Equation~\ref{eq:molfrac} provides just one of the necessary ingredients to compute \mhi for each gas cell. We also require the fraction of gas that is hydrogen, $f_{\rm{H}} = M_{\rm{H}} / M_{\rm{gas}}$, and the fraction of hydrogen that is neutral, $f_{\rm{HI+H}_2} = M_{\rm{HI} + \rm{H}_2} / M_{\rm{H}}$. With these definitions, we can describe the mass of \hi within a gas cell as
\begin{equation} \label{eq:HImass}
    M_{\rm{HI}} = (1 - f_{\rm{mol}}) \times f_{\rm{HI+H}_2} \times f_{\rm{H}} \times M_{\rm{gas}} \,.
\end{equation}
IllustrisTNG provides $f_{\rm{H}}$ and $f_{\rm{HI + H}_2}$ by tracking the abundance of hydrogen, helium and a small network of metals among other physical processes \citep{F-GPhotochem2009, VogelGalFormModel2013, Rahmati2013, CLOUDY2017, PillepichTNGmodel2018}.

However, in star-forming cells, $f_{\rm{HI+H}_2}$ is not computed self-consistently in the IllustrisTNG model \citep{SpringelSFRmodel2003} and is thus treated separately. Star-forming cells are divided into a hot and cold gas phase. The entire hot phase is assumed to be ionized, and any ionization due to local young stars is neglected in the cold phase, yielding $f_{\rm{HI} + \rm{H}_2} = f_{\rm{H}} \rho_{\rm{cold}} / \rho\srm{gas}$. A more detailed discussion of this assumption can also be found in \citet{VogelGalFormModel2013} and \citet{DiemerHI2018}.

To model the final ingredient, $f_{\rm{mol}}$, we must post-process the simulation because IllustrisTNG does not distinguish between molecular and atomic hydrogen. Physically, $f_{\rm{mol}}$ is the result of a balance between H$_2$ production on dust grains and photo-dissociation via Lyman-Werner radiation. Thus $f_{\rm{mol}}$ models require the gas metallicity, gas surface density, and the intensity of incident UV radiation in the Lyman-Werner band as input \citep{DraineGang2011}. We must post-process the simulation to estimate the incident Lyman-Werner UV flux on a particular gas cell because IllustrisTNG does not include radiative transfer. The UV flux on any cell originates from two sources: the cosmological UV background \citep{F-GPhotochem2009} and from nearby star-forming regions. The incident UV flux on gas in galaxies is dominated by nearby stars. The gas in filaments, however, receives significant UV flux from distant sources. However, radiative transfer across the entirety of the simulation volume is impractical. We employ models with two approaches to resolve the filamentary gas issue: \citet{V-NIngred21cm2018}, which neglects stellar UV sources, and \citet{DiemerHI2018}, which neglects gas in filaments. We will briefly describe the relevant portions of each of the two approaches; for more details see the discussed papers.

\citet[][hereafter \citetalias{V-NIngred21cm2018}]{V-NIngred21cm2018} treats star-forming and non-star-forming gas cells separately. \citetalias{V-NIngred21cm2018} assigns $f_{\rm{mol}}=0$ in non-star-forming gas, neglecting the trace amounts of H$_2$. For star-forming gas, $f_{\rm{mol}}$ is calculated using the Krumholz, McKee, \& Tumlinson (KMT) model \citep{KMTMolModelI2008, KMTMolModelII2009}. KMT estimates $f_{\rm{mol}}$ when given properties characterizing the cloud's size, metallicity, and the intensity of the photo-dissociating UV radiation in the Lyman-Werner band \citep{KMTMolModelI2008}. \citetalias{V-NIngred21cm2018} neglects photo-dissociating UV radiation from local star formation, only incorporating background UV estimates from IllustrisTNG itself \citep{F-GPhotochem2009}. \citetalias{V-NIngred21cm2018} do not make any resolution cuts since their model is applied to individual gas cells, although these have a mean mass of $M_{\mathrm{gas}}\gtrsim 9.4 \times 10^5 M_\odot$ \citep{PillepichTNGmodel2018}.

\begin{figure*}
    \centering
    \includegraphics[trim={.8cm 0 11.1cm 0}, clip, width=\linewidth]{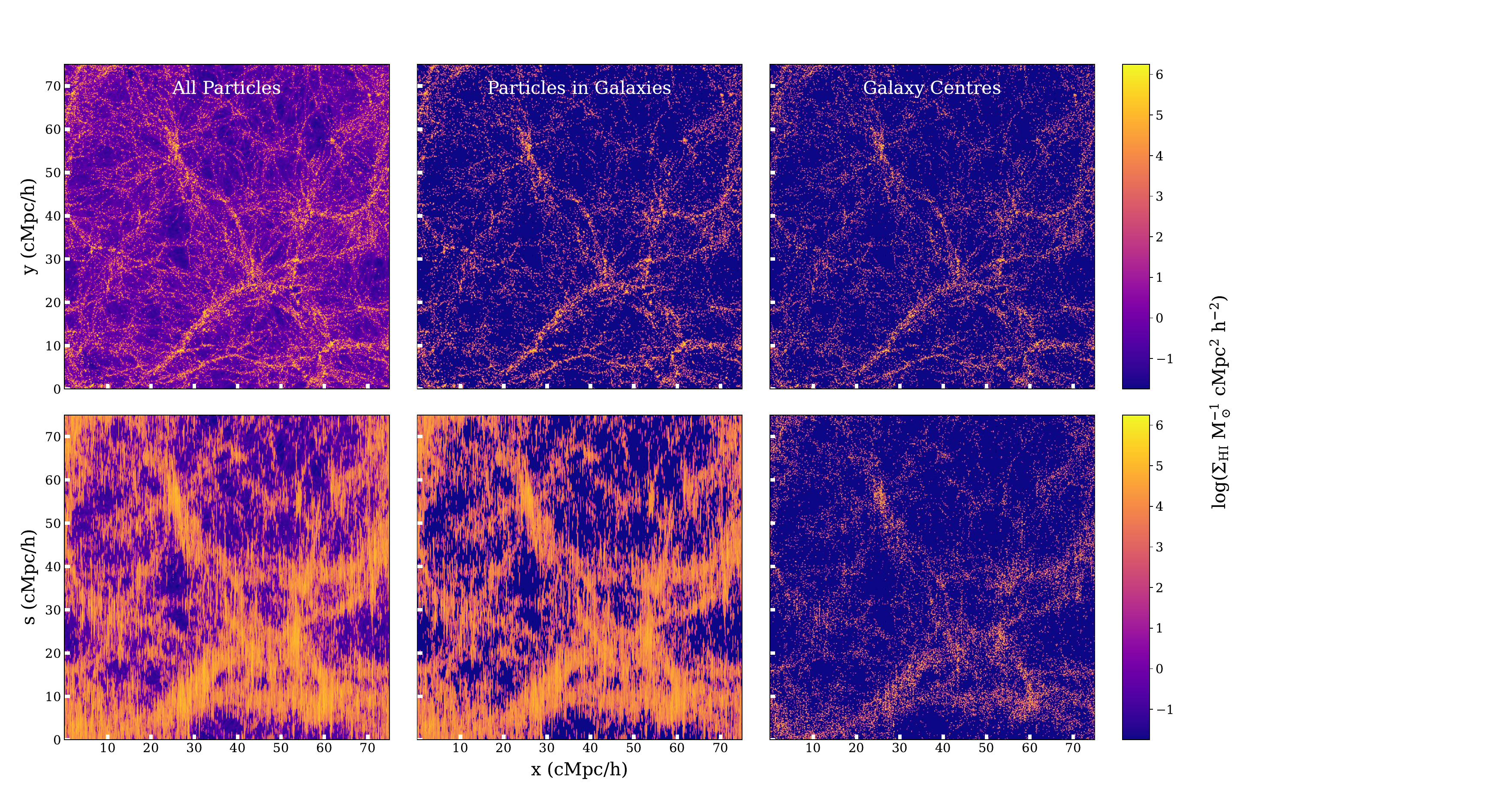}
    \caption{Slices through the IllustrisTNG \hi mass distribution at $z = 0$ in real (top) and redshift space (bottom). Each slice sums 20\% of the simulation volume along the $z$-axis. The redshift-space distributions are created by displacing particles from their positions in real space using their velocities along the line of sight, in this case the $y$-axis. The left column shows the model of \citetalias{V-NIngred21cm2018}, which calculates the \hi in all gas cells but does not account for local UV sources, whereas the remaining distributions account for local UV sources but only account for gas cells within galaxies. The middle and right columns display \hi distributions calculated by \citetalias{DiemerHI2018} with the \citet{GnedinMolModel2014} model, using the positions of individual gas cells and the centre of the host galaxy, respectively. The sizes of the points in the right column scale logarithmically with galaxy \hi mass. In the bottom row, the fingers-of-God effect can be observed in all three cases. In the \textit{Galaxy Centres} case, however, the fingers-of-God manifest more weakly because it removes the contribution of the velocities of the cells within a galaxy along the line of sight.
    }
    \label{fig:slices}
\end{figure*}

\begin{figure}
    \centering
    \includegraphics[width=.65\linewidth, trim={0cm 0 0.1cm 0.1cm}, clip]{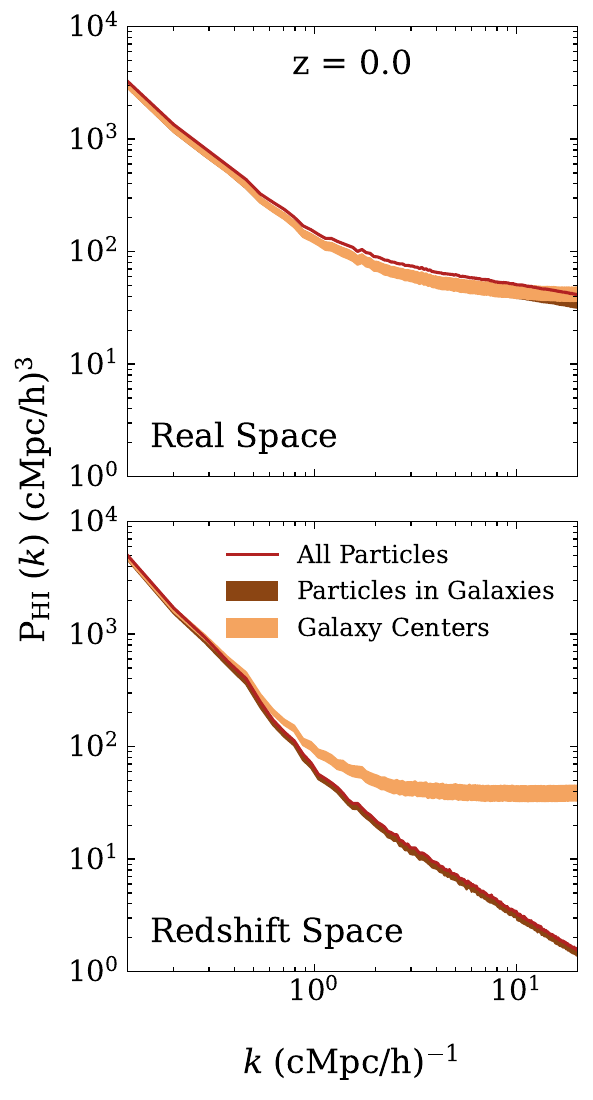}
    \caption{\hi auto power spectra in real (top) and redshift space (bottom) at $z = 0$. The contours delineate the models from \citetalias{DiemerHI2018} that are applied to galaxies as a whole (light brown, nine models) and to the individual particles within galaxies (dark brown, four models). The crimson line represents the distribution from \citetalias{V-NIngred21cm2018}. All distributions are similar in real space, although \textit{All Particles} receives a small boost by including filaments. In redshift space, \textit{Galaxy Centres} diverges from the other auto powers at $k \sim 0.5$ \hmpc. Representing galaxies as points removes any velocity contributions from individual gas cells, tempering the fingers-of-God effect. We represent subsequent \hi power spectra using shaded areas encompassing each model.}
    \label{fig:HI_autopk}
\end{figure}

\begin{figure}
    \centering
    \includegraphics[width=0.65\linewidth]{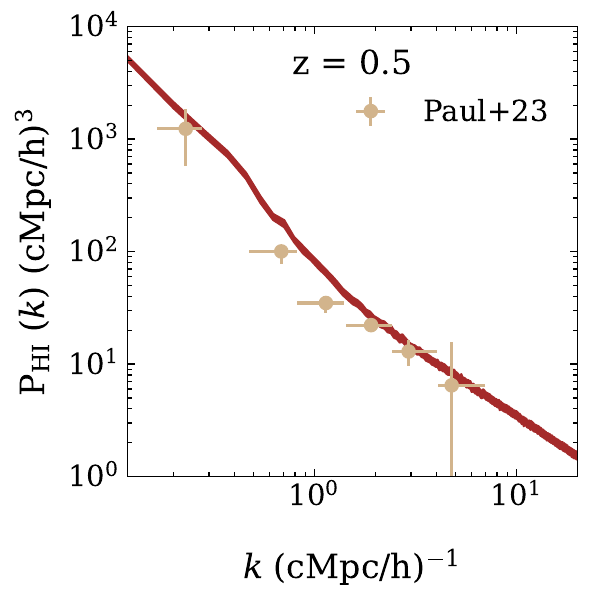}
    \caption{Redshift-space \hi auto power spectra at $z = 0.5$ (brown contour) and the observed auto power spectrum from \citet{Paul2023MEERKATHIauto} at $z \sim 0.44$ (tan points). The contour captures all models from \textit{All Particles} and \textit{Particles in Galaxies}. The agreement between \citet{Paul2023MEERKATHIauto} and IllustrisTNG is encouraging.}
    \label{fig:paul_comp}
\end{figure}
\citet[][hereafter \citetalias{DiemerHI2018}]{DiemerHI2018} utilize five models of three different types to compute $f_{\rm{mol}}$: models based on observed correlations \citep{LeroyMolModel2008}, calibrations with simulations \citep{GnedinMolModel2011, GnedinMolModel2014}, and analytical models \citep{KrumholzMolModel2013,  SternbergMolModel2014}. These models require the same three inputs as the model from \citetalias{V-NIngred21cm2018}: the gas metallicity, density, and the incident UV intensity in the Lyman-Werner band. \citetalias{DiemerHI2018} use the SFR of nearby cells to approximate incident UV from local stellar sources, assuming an optically thin medium (see \citet{gebek23} for comparison to full radiative transfer prescriptions). The models used by \citetalias{DiemerHI2018} are tuned for 2D surface densities, not the 3D densities from the simulation. \citetalias{DiemerHI2018} convert from 3D to 2D in two ways: a cell-by-cell method using the Jean's length and by projecting the galaxy in a face-on orientation onto a 2D grid of pixels (see their section 2.3). Each of the five $f_{\rm{mol}}$ models is then applied to the 2D quantities, except \citet{LeroyMolModel2008} which yields unphysical results when applied cell-by-cell. In total, this procedure results in nine \hi distributions from \citetalias{DiemerHI2018}. These distributions only contain \hi data from galaxies with $M_\star \geq 2 \times 10^8 M_\odot$ or $M_{\mathrm{gas}} \geq 2 \times 10^8 M_\odot$. However, as we will show in Section \ref{subsection:HIcluster}, these resolution limits do not significantly affect the \hi clustering.

In general, the \hi distributions we use agree fairly well with observations, but there are some notable points of tension as described in \citet{DiemerHI2019}. At $z = 0$, IllustrisTNG possesses nearly twice the observed cosmic abundance of \hi. The \hi mass function overall agrees well with observations, but overestimates observed counts at $\mhi \approx 10^9 M_\odot$.

\subsection{Power spectra}
\label{subsection:calc_pk}

We compute the power spectra of the overdensity $\delta(\boldsymbol{x}) = \rho(\boldsymbol{x}) / \overline{\rho} - 1$ with
\begin{equation}
\label{eq:pk}
    \langle \Tilde{\delta}_i(\boldsymbol{k}) \Tilde{\delta}_j(\boldsymbol{k}') \rangle = (2\pi)^3 P_{i \times j}(k) \delta_D^3(\boldsymbol{k} - \boldsymbol{k}') \,.
\end{equation}
$P_{i \times j}(k)$ is the power spectrum and $\boldsymbol{k}$ is the wavenumber, with bold denoting a vector. $\delta_D$ is the Dirac delta function. $\Tilde{\delta_i}$ represents the Fourier transform of the overdensity from position-space to $k$-space. If $i=j$ in the above equation, $P_{i \times j}(k)$ is called an auto power spectrum, which will be denoted with one population $P_i(k)$. Otherwise, it is called a cross-power spectrum.

The halo occupation distribution \citep{PeacockHOD2000, BerlindHOD2002, KravstovHOD2004, TinkerHOD2005, ZhengHOD2007, hadzhiyskaHOD2020} provides a useful analytical framework for understanding power spectra. We can represent the power spectrum as a sum of contributions of inter- and intra-halo galaxy pairs,
\begin{equation} \label{eq:hod}
    P(k) = P^{2\rm{h}}(k) + P^{1\rm{h}}(k) + P^{\rm{SN}} \,.
\end{equation}
The two-halo term reflects large-scale structure and the one-halo term structure within haloes. The shot noise term is a constant determined by the size of the sample used to measure the clustering. In the following sections, we will often refer to this framework to guide our analysis of the power spectra.

We consider power spectra in both real and redshift space. Matter is placed in redshift space $\boldsymbol{s}$ by displacing real-space positions along an arbitrarily chosen line-of-sight using their velocities, 
\begin{equation}
    \label{eq:rss}
    \boldsymbol{s} = \boldsymbol{x} + \frac{1 + z}{H(z)} \boldsymbol{v_\parallel} (\boldsymbol{r}) \,,
\end{equation}
where $H(z)$ is the Hubble parameter, $\boldsymbol{x}$ is position in real space, and $\boldsymbol{v_\parallel}$ is the velocity parallel to the chosen line of sight. The resulting real- and redshift-space matter distributions are placed in a $800^3$ grid with bin lengths of $\approx$ 95 \units{\textit{h}}{-1} ckpc, binned using a Cloud-In-Cell (CIC) interpolation scheme. The grids are then used as input for the power spectrum calculation routine provided by the Python library \textsc{Pylians} \citep{Pylians}. Our results are converged with grid resolution at all relevant scales (online figures).

We can quantify the ``faithfulness'' of a matter tracer using the bias $b_{ij} (k) = \Tilde{\delta_i} (\boldsymbol{k}) / \Tilde{\delta_j} (\boldsymbol{k})$, where $i$ and $j$ represent two different populations. For this paper, we are exclusively interested in the bias of matter tracers with respect to the full mass distribution, so we always take $j$ to mean ``all matter'' and express the bias as $b_i$. We calculate the bias as
\begin{equation}
\label{eq:bias}
    P_i (k) = b_i^2 (k) P_{\rm{m}} (k) \,
\end{equation}
where $P_{\rm{m}}$ is the matter power spectrum, and $i$ represents some chosen matter tracer such as \hi or galaxies.

To measure the strength of the relationship between two samples, we use the correlation coefficient
\begin{equation}
\label{eq:corr_coef}
    r_{i-j} = \frac{P_{i \times j} (k)}{\sqrt{P_i (k) P_j (k)}} \,.
\end{equation}
 $r_{i-j}$ takes a value of zero for completely random distributions and approaches unity for entirely dependent samples. We distinguish the correlation coefficient from the position vector by denoting the vector $\boldsymbol{r}$ using a bold letter and the length scale $R$ with a capital.

 \section{Results}
\label{section:results}

In this section, we characterize the clustering of \hi, blue, and red galaxies. In Section~\ref{subsection:HIcluster}, we show that the \hi distribution is not particularly sensitive to the post-processing of the simulations. We analyse how \hi clusters with different galaxy samples separated by colour in Section~\ref{subsection:HIcolor} and examine their scale- and time-dependence in Section~\ref{sub:zevo}. We study how the cross-power spectra change as a function of $M_\star$ in Section~\ref{subsection:stbinth} and \mhi in Section~\ref{subsection:hibinth}. In Appendix~\ref{appendix:convergence}, we show how simulation resolution affects these results.

\subsection{H\textsc{i} auto power spectra} \label{subsection:HIcluster}

\begin{figure}
    \centering
    \includegraphics[width=\linewidth, trim={0.25cm 0.05cm 2.5cm 10.5cm}, clip]{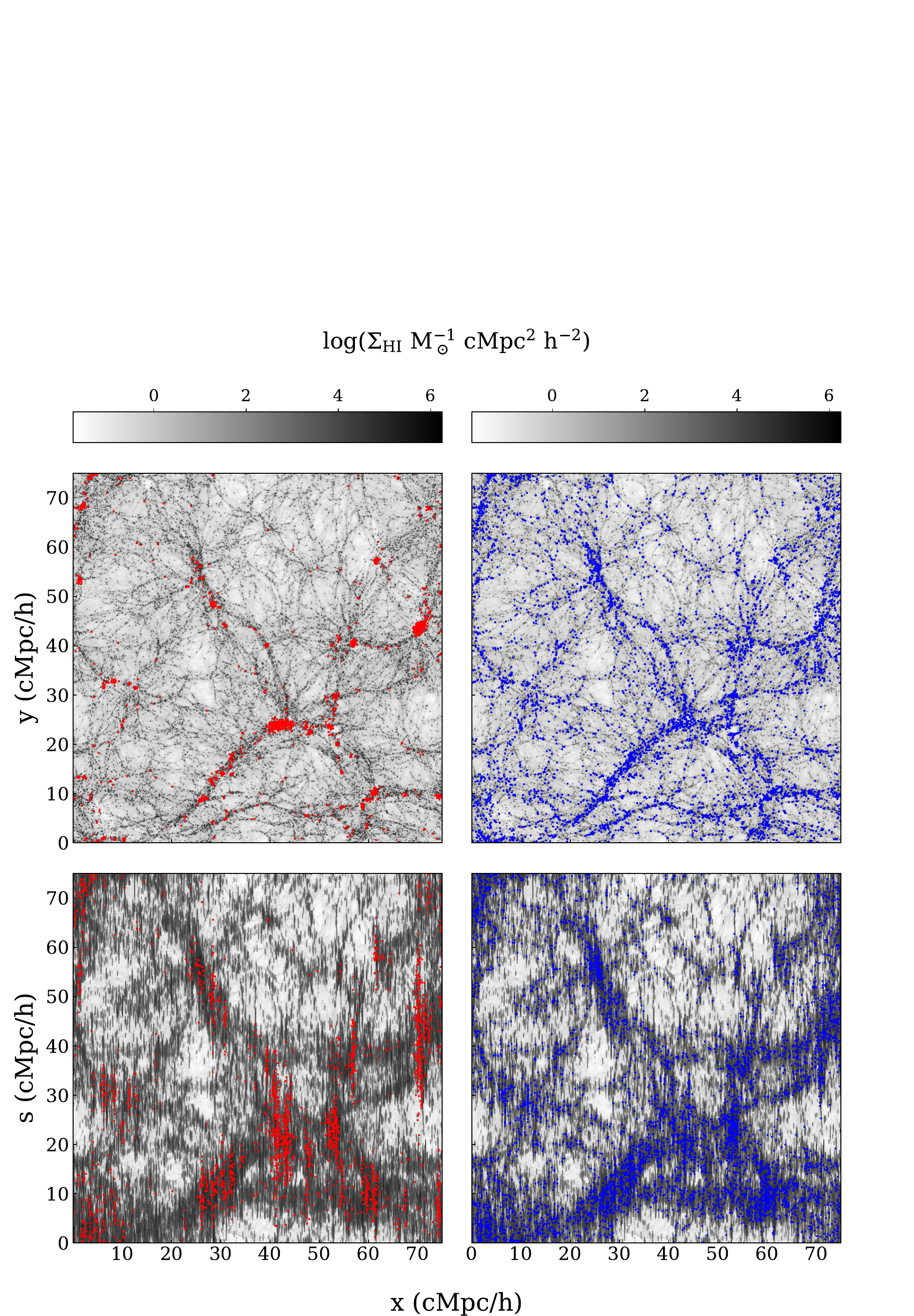}
    \caption{Slices of the red (left) and blue (right) galaxy distributions overlaid with the \hi distribution from \citetalias{V-NIngred21cm2018}. The slices show TNG100 at $z = 0$, summing 20\% of the $z$-plane centred in the middle of $z$-axis, in both real (top) and redshift (bottom) space. Matter was placed in redshift space by projecting its velocities along the line of sight, in this case the $y$-axis. Red galaxies are clearly more clustered, grouping heavily around the largest haloes, whereas the blue galaxies better trace out the \hi distribution across the box.}
    \label{fig:rvb_slice}
\end{figure}

\begin{figure*}
    \centering
    \includegraphics[width=\linewidth]{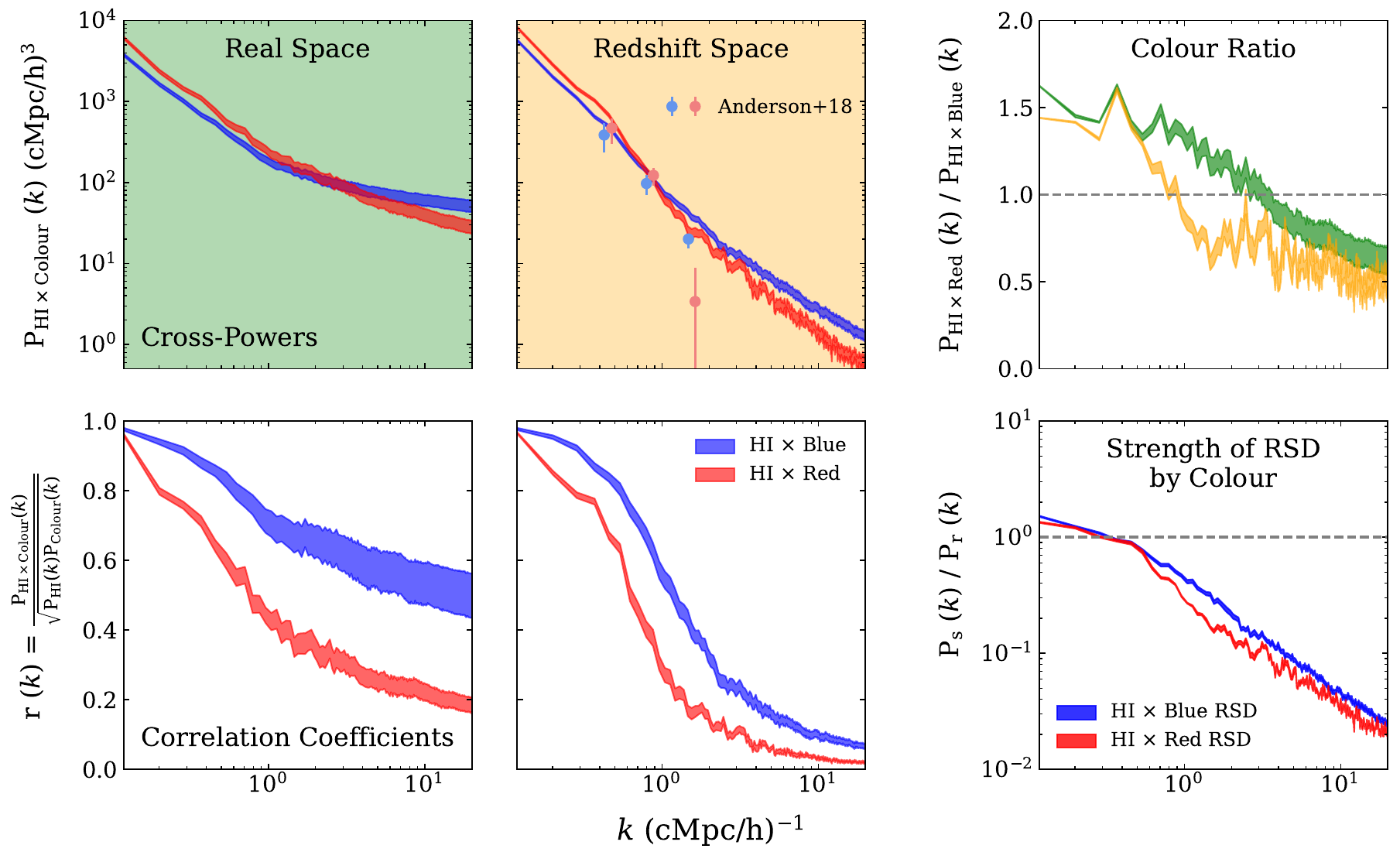}
    \caption{Cross-power spectra (top) and correlation coefficients (bottom) for \hir and \hib at $z = 0$ in real space (left, green) and redshift space (centre, yellow), shown as contours that enclose the \hi models. The cross-power spectra are calculated with respect to the total \hi distribution, not only the \hi in galaxies of the respective colour. We compare the observational data from \citet{AndersonHI-color2018} to the redshift-space power spectra, displayed as points. For clarity, red points are offset in $k$. The point at which the cross-power spectra intersect is more discernible in the ratio \hir over \hib (top right) for both real (green, $k\srm{eq} \approx 3.25$ \hmpc) and redshift space (yellow, $k\srm{eq} \approx 0.9$ \hmpc). We compare the ratio of redshift-space over real-space power spectra for \hir and \hib (bottom right). At small scales, \hir is suppressed more than \hib because FoG are stronger in red galaxies. This adds a secondary colour-dependency, which pushes the intersection between the cross-power spectra out to $\sim 10.4$ Mpc in redshift space. The correlation coefficients confirm that \hi is more tightly linked to blue galaxies than red at all scales in both real and redshift space.}
    \label{fig:rvb}
\end{figure*}

Before we proceed, we must ascertain if our results are sensitive to the post-processing models used to calculate \hi distributions in IllustrisTNG. We split the \hi models into three groups: \textit{Galaxy Centres}, \textit{Particles in Galaxies}, and \textit{All Particles}. For models in \textit{Galaxy Centres}, we assign all the \hi in a galaxy to its centre for all nine models from \citetalias{DiemerHI2018}. The four cell-by-cell models from \citetalias{DiemerHI2018} are also included in the next group, \textit{Particles in Galaxies}, where we instead assign \hi to the position of each individual host cell. Gas cells outside of galaxies are excluded in both cases (Section~\ref{subsection:phase}). For the final group, \textit{All Particles}, we apply the one model from \citetalias{V-NIngred21cm2018} to every gas cell in the simulation. In this section, we compare the \hi clustering for the \hi models in and amongst these three groups.

We show slices through the real-space \hi distributions in the top row of Fig.~\ref{fig:slices} to provide insight into the \hi auto power spectra in the top panel of Fig.~\ref{fig:HI_autopk}. The filaments present in \textit{All Particles} (top left in Fig.~\ref{fig:slices}) but missing in \textit{Galaxy Centres} and \textit{Particles in Galaxies} (top middle and right, respectively) reflect the trade-off described in Section~\ref{subsection:phase} -- namely, the neglect of local UV sources in \citetalias{V-NIngred21cm2018} and of filaments and low-mass galaxies in \citetalias{DiemerHI2018}. These galaxies below the mass thresholds from \citetalias{DiemerHI2018} (see Section~\ref{subsection:phase}) and the filaments that connect galaxies supply some small mass-weighted power, boosting the \hi power spectrum from \textit{All Particles} slightly on all scales in Fig.~\ref{fig:HI_autopk}. On small scales, \textit{Galaxy Centres} diverges from \textit{Particles in Galaxies}, as representing a galaxy's \hi as a point removes any clustering within galaxies. For all power spectra presented in this work, we exclude wavenumbers above $k \approx 20$ \hmpc due to aliasing issues near the Nyquist frequency ($k_{\rm{Nyq}} = 33$ \hmpc).

Measurements of \hi, however, take place in redshift space. Line-of-sight velocities distort the real-space power spectra primarily in two ways, called redshift-space distortions (RSD): the Kaiser effect \citep{Kaiser1987} and fingers-of-God \citep[FoG,][]{FOG1972}. The Kaiser effect enhances each redshift-space auto power spectrum (bottom panel of Fig.~\ref{fig:HI_autopk}) on large scales, as groups of galaxies moving coherently appear closer in redshift space. The second RSD, FoG, manifests due to the velocity dispersions within a halo, smearing the distribution on small scales into the columns (``fingers'') along the line of sight (bottom left and centre panels of Fig.~\ref{fig:slices}). FoG suppress the redshift-space power spectrum compared to the real-space counterparts for each \hi group (bottom of Fig.~\ref{fig:HI_autopk}), although the magnitude of the suppression varies between groups.

FoG manifest more weakly in \textit{Galaxy Centres} than in \textit{Particles in Galaxies} and \textit{All Particles}. The strength of FoG in each \hi group is dependent on its velocity dispersion \citep{Juszkiewicz2000PVD}. The velocity dispersion for \textit{Particles in Galaxies} and \textit{All Particles} receives contributions from the velocity dispersion of galaxies themselves \textit{and} their constituent particles \citep{zhang2020}. By collapsing \hi to the centre of each galaxy, \textit{Galaxy Centres} removes contributions from the intrinsic velocity dispersion of the particles within a galaxy, softening the net FoG in the \textit{Galaxy Centres} case. As a result, the redshift-space \hi auto power for \textit{Galaxy Centres} breaks away from the other two groups at $k \gtrsim 0.5$ \hmpc, before approaching a constant value due to shot noise at $k \sim 2$ \hmpc.

We compare the redshift-space \hi auto power spectra from \textit{All Particles} and \textit{Particles in Galaxies} to observations from \citep{Paul2023MEERKATHIauto} in Fig. \ref{fig:paul_comp}, finding strong agreement between the two at small and large scales. At intermediate scales ($k \sim 0.8$\hmpc), IllustrisTNG slightly overpredicts observed results. However, the strong agreement between the two is reassuring, indicating that the spatial distribution of \hi in IllustrisTNG is realistic at $z = 0.5$. We exclude the second $k$ bin from \citet{Paul2023MEERKATHIauto}, as the measurement is dominated by systematics from small baselines (see their appendix A).

In following sections, we represent \hi power spectra as shaded areas encompassing each of the \hi models as in Fig. \ref{fig:paul_comp}, treating the contours as systematic uncertainties due to \hi post-processing. However, we continue to exclude \textit{Galaxy Centres} from redshift-space power spectra since their FoG are artificially suppressed, which is unrelated to \hi post-processing.

\subsection{H\textsc{i}-Galaxy cross-power spectra}
\label{subsection:HIcolor}

Previous works have established the environmental dependence of a galaxy's gas abundance, finding that cluster members have significantly lower gas fractions \citep{GiovanelliGasDefClus1985, Solanes2002, Brown2021VERTICO}. We can measure the effect of this environmental dependence on larger scales by computing cross-power spectra between \hi and blue galaxies (\hib) and \hi and red galaxies (\hir). In Section~\ref{HIColor:hicreal} we focus on their scale- and colour-dependence in real space. In Section~\ref{HIColor:hicred}, we study the impact of RSDs on those relationships. We study the impact of varying colour cuts on our cross-power spectra in Appendix \ref{appendix:color_cut}, finding that the clustering on large scales ($k \lesssim 1$ \hmpc) does not depend on colour cut. On smaller scales where the sensitivity increases, we instead focus on trends.

\subsubsection{Real space} \label{HIColor:hicreal}

The top row of Fig.~\ref{fig:rvb_slice} shows the distribution of blue and red galaxies overlaid on \hi in real space. As mentioned in Section \ref{section:methods}, we use galaxies with $M_\star \geq 2 \times 10^8$ for all results derived from galaxy populations. Red galaxies (top left panel) are concentrated in the densest regions while blue galaxies (top right) are broadly distributed. Massive, older, and more clustered haloes tend to host red galaxies, leading to colour-dependent clustering and a larger bias (equation~\ref{eq:bias}) with respect to matter for red galaxies than blue galaxies \citep{GaoAgeClustering2005, ZehaviLumColCluster2005, WechslerAssemblyBias2006, SpringelTNG2018}. The intrinsic clustering strength of red galaxies manifests in the cross-power spectra shown in Fig.~\ref{fig:rvb} via their larger bias. Mathematically, we can express the cross-power spectra as
\begin{equation}
\label{eq:crosspk}
    P_{\mathrm{HI} \times \mathrm{Colour}} (k) = b\srm{HI} (k) b\srm{Colour} (k) r\srm{HI-Colour} (k) P\srm{m} (k) 
\end{equation} 
to describe how they relate to the bias $b\srm{Colour}$, correlation coefficient $r\srm{HI-Colour}$, and matter power spectrum $P\srm{m}$. Equation~\ref{eq:crosspk} is useful for understanding the galaxy properties responsible for the relative strengths of \hib and \hir: their inherent clustering strength, represented by their bias $b\srm{Colour}$, and their spatial relationship with \hi, represented by the correlation coefficient $r\srm{HI-Colour}$ (equation~\ref{eq:corr_coef}).

\hir is greater than \hib on large scales (Fig.~\ref{fig:rvb}) because red galaxies cluster more strongly ($b\srm{Red} > b\srm{Blue}$). However, this trend is counteracted by the weaker spatial connection between red galaxies and \hi ($r\srm{HI-Red} < r\srm{HI-Blue}$, bottom row). The disparity between the correlation coefficients is small on large scales ($r\srm{HI-Red} \approx r\srm{HI-Blue} \approx 1$) because the clustering of \hi is colour-independent, in the sense that its distribution is governed mostly by linear growth rather than small-scale effects. 
However, \hi tends to be suppressed in the massive haloes hosting red galaxies, reducing $r\srm{HI-Red}$ faster than $r\srm{HI-Blue}$ when approaching small scales. The disparity between the correlation coefficients grows such that on small scales the \hib cross-power spectrum is greater than \hir. We describe the clustering of \hi on these scales as ``colour-dependent'' since the spatial relationship between \hi and galaxies governs the relative strengths of each cross-power spectra more than the galaxy population's inherent clustering. 

In terms of equation \ref{eq:crosspk}, we can call those scales colour-independent when \hir is greater than \hib, as $P_{\mathrm{HI} \times \mathrm{Red}} / P_{\mathrm{HI} \times \mathrm{Blue}} > 1$ implies $b\srm{Red}/b\srm{Blue} > r\srm{HI-Blue} / r\srm{HI-Red}$ and therefore the intrinsic clustering of the galaxy population dictates the relative strengths of the cross-power spectra on large scales. On the other hand, colour-dependent scales occur when \hib is greater than \hir, as $P_{\mathrm{HI} \times \mathrm{Red}} / P_{\mathrm{HI} \times \mathrm{Blue}} < 1$ implies $b\srm{Red}/b\srm{Blue} < r\srm{HI-Blue} / r\srm{HI-Red}$, showing that the galaxy population's spatial relationship with \hi is largely responsible for their relative strengths. We roughly interpret the scale at which the intersection between \hir and \hib occurs ($k\srm{eq}$) as the transition between the colour-independent and -dependent regimes, but emphasize that $k\srm{eq}$ is not a sharp transition. 

The intersection is easily identifiable when the ratio of \hir over \hib (top right panel of Fig.~\ref{fig:rvb}) falls below unity at $k_{\rm{eq}} \approx 3.25$ \hmpc ($\req = 2\pi / k_{\rm{eq}} \approx 2.8$ Mpc). The scale at which the \hi distribution becomes significantly more colour-dependent reflects the findings of previous work on the concept of ``galaxy conformity''. Galaxies within $\sim 4$ Mpc of a larger red galaxy tend to also be red \citep{Kauffmann2013GalConformity}, indicating that a galaxy's large-scale environment impacts its star-formation \citep{hearin2016physical, AyromlouConform2023}. Given the link between \hi and star-formation \citep{BigielSFR-HI2008}, it is reasonable that the \hi content of galaxies would also be suppressed within 4 Mpc of a large red galaxy. This crossover also matches the $z \approx 0$ cross-correlation between SDSS galaxies \citep{SDSS2000} and ALFALFA \hi maps \citep{ALFALFA2005, ALFALFA2011} from \citet{BigPapa2013}, where they find that \hi abundance is reduced within 3 Mpc ($k \sim 3.14$ \hmpc) of a red galaxy. The agreement between IllustrisTNG and observations is encouraging.

\subsubsection{Redshift space}\label{HIColor:hicred}

Fig. \ref{fig:rvb_slice} shows slices through the redshift-space distributions of red and blue galaxies overlaid on \hi (bottom row), where the FoG in red galaxies stretch over longer distances than in blue galaxies \citep{li2006dependence}. The massive haloes that host red galaxies possess deep potential wells, causing a large velocity dispersion in the member galaxies that can stretch FoG for nearly half the length of the box. This colour-dependency in the strength of FoG alters the comparison between \hib and \hir in the redshift-space cross-power spectra, which are provided in Fig.~\ref{fig:rvb} (top middle panel).

We quantify the strength of RSDs on the power spectra with the ratio of real- and redshift-space power spectra (bottom right panel of Fig.~\ref{fig:rvb}). At large scales, the redshift-space power spectra are slightly greater than their real-space counterparts until $k \sim 0.4$ \hmpc. The boost arises from the Kaiser effect (Section~\ref{subsection:HIcluster}) and manifests similarly in both \hib and \hir. Although both \hib and \hir will approach the Kaiser limit at sufficiently large scales, it is unclear if they will approach the limit similarly, and it is difficult to extrapolate from the small $k$ regime that probes the Kaiser effect. At small scales, FoG dominate the distortion of redshift-space power spectra, reducing \hib and \hir significantly within $k > 0.4$ \hmpc. FoG suppress \hir more strongly than \hib, with the two diverging at $k \sim 0.5$ \hmpc. The stronger FoG in \hir introduce a secondary colour-dependent effect in the redshift-space cross-power spectra. Consequently, the redshift-space colour ratio (top right panel of Fig.~\ref{fig:rvb}) curves downward much more rapidly at $k \approx 0.4$ \hmpc{} and reaches unity at a much larger scale in redshift-space ($\req \approx 10.4$ Mpc) than in real space ($\req \approx 2.8$ Mpc). At $k > 1$ \hmpc, random particle velocities emerge as noise in the redshift-space power spectra---this effect is small and does not alter any conclusions made in this paper.

Below the redshift-space cross-power spectra in Fig.~\ref{fig:rvb}, we showcase the $z = 0$ redshift-space correlation coefficients. Similarly to their real-space counterparts, \hib has a greater correlation coefficient than \hir at all scales. However, both \hir and \hib decrease much more sharply at $k \sim 1$ \hmpc than in real-space, with their redshift-space correlation coefficients reaching $\sim 0.1$ on the smallest scales. We speculate that the spatial disconnect may arise from differences in velocity dispersions between \hi and galaxies. \hi possesses intrinsic velocity dispersion within galaxies \citep{zhang2020}, whereas galaxies only experience pair-wise velocity dispersion. We reserve further analysis for the following paper (Osinga et al., in prep).

The power spectra shown in Fig.~\ref{fig:rvb} agree with $z \sim 0$ observations. \citet{AndersonHI-color2018} computed cross-correlations between 2df galaxies \citep{2df2001} and Parkes \hi maps \citep{Parkes1996}, shown as points in the top centre panel of Fig.~\ref{fig:rvb}. Our results match in the range $0.3 < k < 1.5$ \hmpc{}, with both \hib{} and \hir{} within their statistical uncertainties. We both find that \hir{} is greater on large scales but drops beneath \hib{} at smaller scales, intersecting within $0.8 < k < 1.5$ \hmpc{} range. \citet{AndersonHI-color2018} measure \hi suppression around red galaxies to much larger scales than \citet{BigPapa2013} ($\sim 10$ Mpc compared to $\sim 3$ Mpc) because of RSDs. \citet{BigPapa2013} employ a projected correlation function that measures clustering perpendicular to the line of sight, removing RSD effects. At larger scales $k \sim 0.1-0.3$ \hmpc, \citet{AndersonHI-color2018} find anti-correlations between \hi and galaxies, which we exclude from Fig.~\ref{fig:rvb} due to their large uncertainties. Interestingly, other intensity mapping experiments find subdued (although still positive) \hi clustering at similar scales even at other redshifts \citep{WolzeBOSSGBT2021, Paul2023MEERKATHIauto}, however we find no evidence for such behaviour in IllustrisTNG.

\subsection{Redshift evolution} 
\label{sub:zevo}

\begin{figure}
    \centering
    \includegraphics[width = \linewidth]{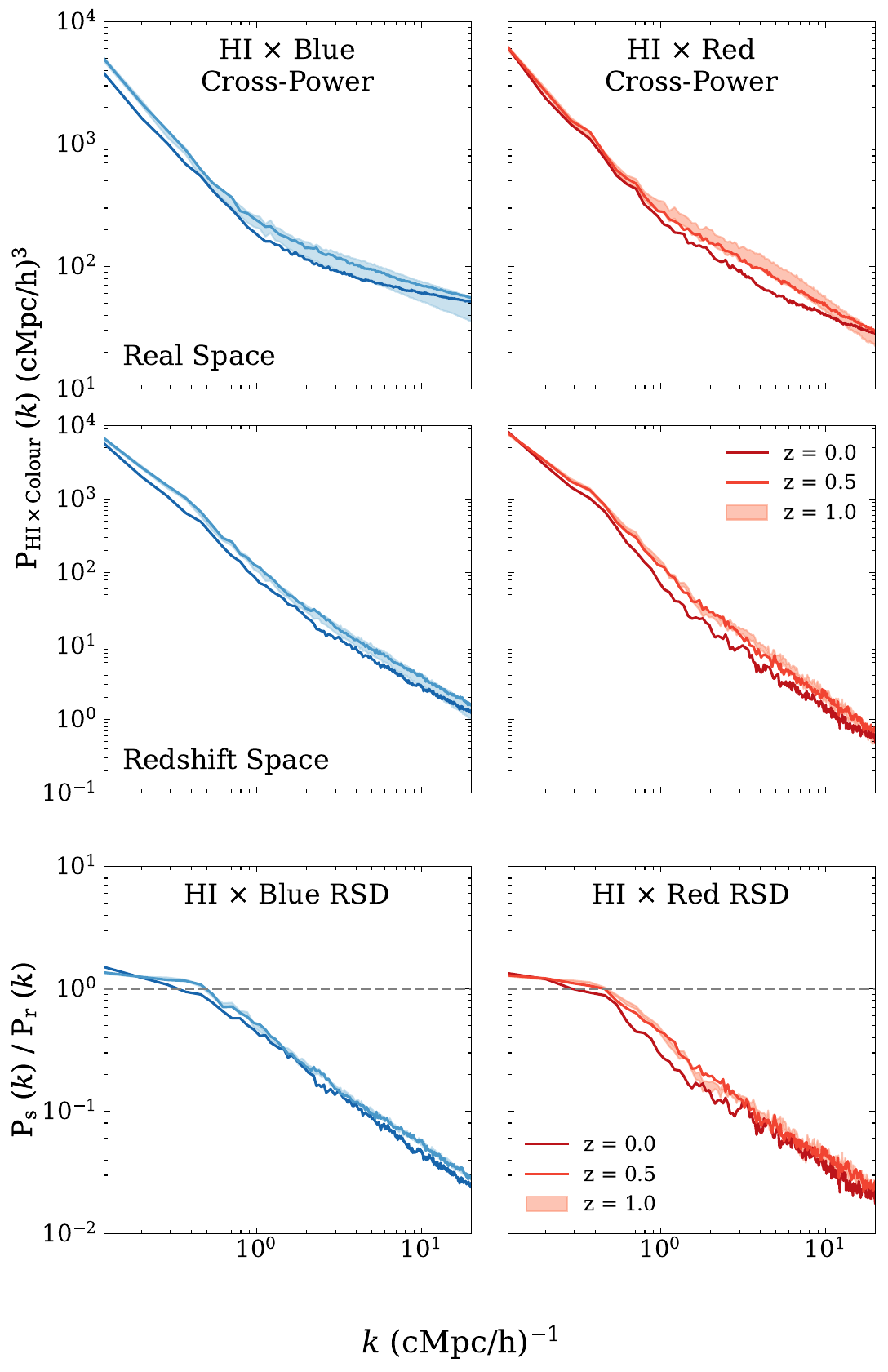}
    \caption{Redshift evolution of \hib (left) and \hir (right) in real (top) and redshift (centre) space for $z = 0, 0.5$, and 1 (darkest to lightest colour). The bottom row shows the redshift evolution of the ratio of redshift-space over real-space power spectra, measuring the strength of RSDs. For clarity, we only plot the contour with all \hi models for $z = 1$ and plot the others as a line representing the contour's centre---the width of the contours does not change significantly across redshift. In real and redshift space, both cross-power spectra either change little or \textit{decrease} with time. These trends seem to conflict with the picture of structure growth but occur due to colour transitions and gas loss (see text). For each colour, RSDs strengthen slightly with time, amplifying the loss of clustering at later times in redshift space as compared to real space.}
    \label{fig:rsd_zevo}
\end{figure}

\begin{figure}
    \centering
    \includegraphics[width=\linewidth, trim = {0 0.3cm 0 0}, clip]{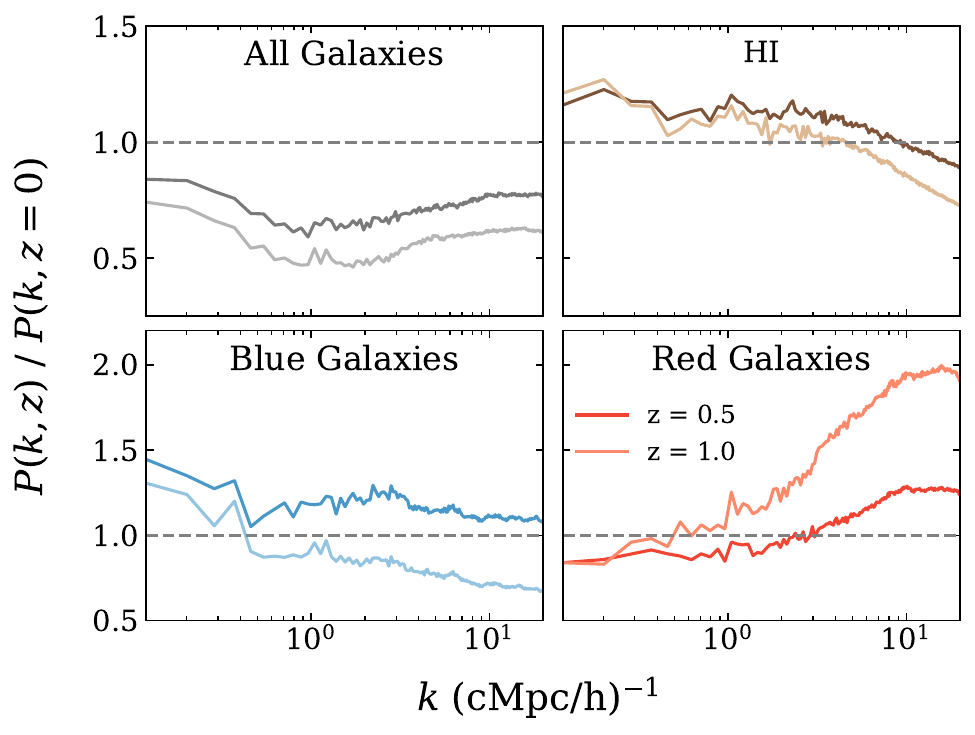}
    \caption{Redshift evolution of the auto power spectra of \hi (top right), blue galaxies (bottom left), red galaxies (bottom right), and the blue and red galaxy population combined (All Galaxies, top left). All power spectra are in real space and are normalized by the $z = 0$ auto power spectra. Darker colours indicate results closer to $z = 0$. The \hi auto powers are displayed using the centre of the contour containing all the \hi models. The only auto power that consistently increases with time is the all-galaxy auto power spectrum. This implies that the decrease in the clustering of red and blue galaxies with time arises due to colour transitions.}
    \label{fig:auto_zevo}
\end{figure}

Assuming that gravity is the dominant influence on the time evolution of power spectra, we expect the clustering of matter to increase linearly on large scales with respect to the growth factor and non-linearly on small scales \citep{Fry1996biaspassive, Dodelson2020}. However, the behaviour of \hib and \hir in Fig.~\ref{fig:rsd_zevo} deviates from those expectations. From $z = 1$ to $z = 0.5$, \hib{} and \hir{} experience negligible changes and from $z = 0.5$ to $z = 0$ their power spectra even \textit{diminish} with time. This discrepancy prompts us to investigate the various processes responsible for the time evolution of each cross-power spectrum. Since cross-power spectra conflate changes in the member distributions and their spatial relationship (Section~\ref{subsection:HIcolor}), we analyse the auto power spectra of the member distributions first in Fig.~\ref{fig:auto_zevo} and then apply those insights to the \hi-galaxy cross-power spectra.

Throughout this section, we examine three interconnected processes that shape the redshift evolution of power spectra: galaxy quenching, gas loss, and colour transitions. While these processes are certainly linked, it is important to clarify their precise meanings in our analysis. We employ the term ``quenching'' to describe galaxies leaving the star-formation main sequence \citep{Daddi2007, Noeske2007, DonnariUVJTNG2019}. ``Gas loss'' refers to processes that strictly affect \hi clustering by reducing a galaxy's gas reservoirs, regardless of when the gas loss occurs relative to quenching. A ``colour transition'' denotes a galaxy's transition from blue to red along the \gr axis, again independent of quenching. Galaxy quenching, per these definitions, is largely not responsible for the redshift evolution of the power spectra, but it is closely associated with other two processes that do contribute to this trend. For the remainder of the paper, we refer to quenching, gas loss, and colour transitions with these definitions in mind.

The bottom row of Fig.~\ref{fig:auto_zevo} shows that both the blue and red galaxy auto power spectra decrease with time from $z = 1$ to $z = 0$, \textit{despite} the all-galaxies auto power spectrum (top left panel) continuing to grow. This discrepancy implies that the decreasing auto power spectra for both blue and red galaxies is an artefact of making a colour cut. Some fraction of galaxies transition from blue to red between redshifts. These transitioning galaxies tend to be the ``reddest,'' oldest, and most clustered members of the blue population. As a result, the average clustering of the remaining blue galaxies decreases upon losing their most clustered component (bottom left panel). Simultaneously, transitioning galaxies join the red population as its \textit{least} clustered members, thereby reducing the average clustering of the red galaxies and decreasing their power spectrum (bottom right panel). The transitioning population at $z \leq 1$ is large enough to offset structure growth, particularly for red galaxies between $z = 1$ and $z = 0.5$ and blue galaxies between $z = 0.5$ and $z = 0$ when their smaller populations increase sensitivity to transitioning galaxies (Fig.~\ref{fig:gr-stmass}). Importantly, it should be noted that this is not a result of moving the \gr cut with time (Section \ref{section:methods}); these trends are amplified if the colour cut remains constant.

As of yet, the effect of colour transitions on the clustering evolution of \hi and galaxies has not yet been properly understood. Previous observational studies on this subject focus on massive red galaxies \citep{zehavi2005LRGzevo, White2007, Guo2013SDSS}, as it is challenging to obtain complete samples of blue galaxies at higher redshifts \citep{Wang2021} and down to the small stellar masses we study here. These studies have found that massive red galaxy clustering grows at a slower rate than all-galaxy clustering over $0 \lesssim z \lesssim 1$, a phenomenon attributed to mergers and disruptions. However, our findings suggest that colour transitions are largely responsible for suppressing the clustering growth of blue and red galaxies. Blue galaxies merge at a slower rate than red galaxies \citep{linmerger2008, Darg2010}, meaning that mergers and disruptions are unlikely to induce similar effects on the redshift evolution of the blue galaxy auto power spectrum (bottom left panel of Fig.~\ref{fig:auto_zevo}). We will further discuss the contributions from mergers, disruptions, and colour transitions on the time evolution of the power spectra at different scales in Section~\ref{section:discussion}.

In addition to the changes in the clustering of blue and red galaxies, the time evolution of \hib and \hir is influenced by the behaviour of \hi clustering. The upper right panel of Fig.~\ref{fig:auto_zevo} shows that the \hi auto power spectrum increases between $z = 1$ and $z = 0.5$ before decreasing between $z = 0.5$ and $z = 0$. Given the connection between blue galaxies and \hi, we speculate that the gas loss associated with colour transitions plays a significant role in the loss of structure from $z = 0.5$ to $z = 0$. Notably, the cosmic abundance of \hi increases from $z = 1$ to $z = 0$ in IllustrisTNG \citep{V-NIngred21cm2018, DiemerHI2019}, which would seem to contradict our finding that the clustering decreases with time. This may be because of changes in the clustering of \hi as a function of halo mass, however further investigations of this conflict are left to future work.

The redshift-space power spectra evolve with time similarly to their real-space counterparts. The strength of this trend is amplified in redshift space by the evolution of RSDs (bottom row of Fig. \ref{fig:rsd_zevo}). For both colours, the RSDs suppress the redshift-space power spectra more strongly at later redshifts, albeit the relative weak evolution of the RSDs overall. Although not explicitly visible in Fig.~\ref{fig:rsd_zevo}, the redshift-space intersection between \hib and \hir occurs at $\req \approx 10.4, 5.4$, and 3.8 Mpc at $z = 0$, 0.5, and 1, respectively. These intersections at $z = 0$ and $z = 1$ are comparable to those reported by \citet{AndersonHI-color2018} and \citet{WolzeBOSSGBT2021}, respectively. We refrain from direct comparisons between IllustrisTNG and \citet{WolzeBOSSGBT2021}, since the overlapping scales between the two are heavily affected by beam effects. Accounting for this requires mock-observing the IllustrisTNG data, which we leave for future work.

\begin{figure*}
    \centering
    \includegraphics[width = .75\linewidth]{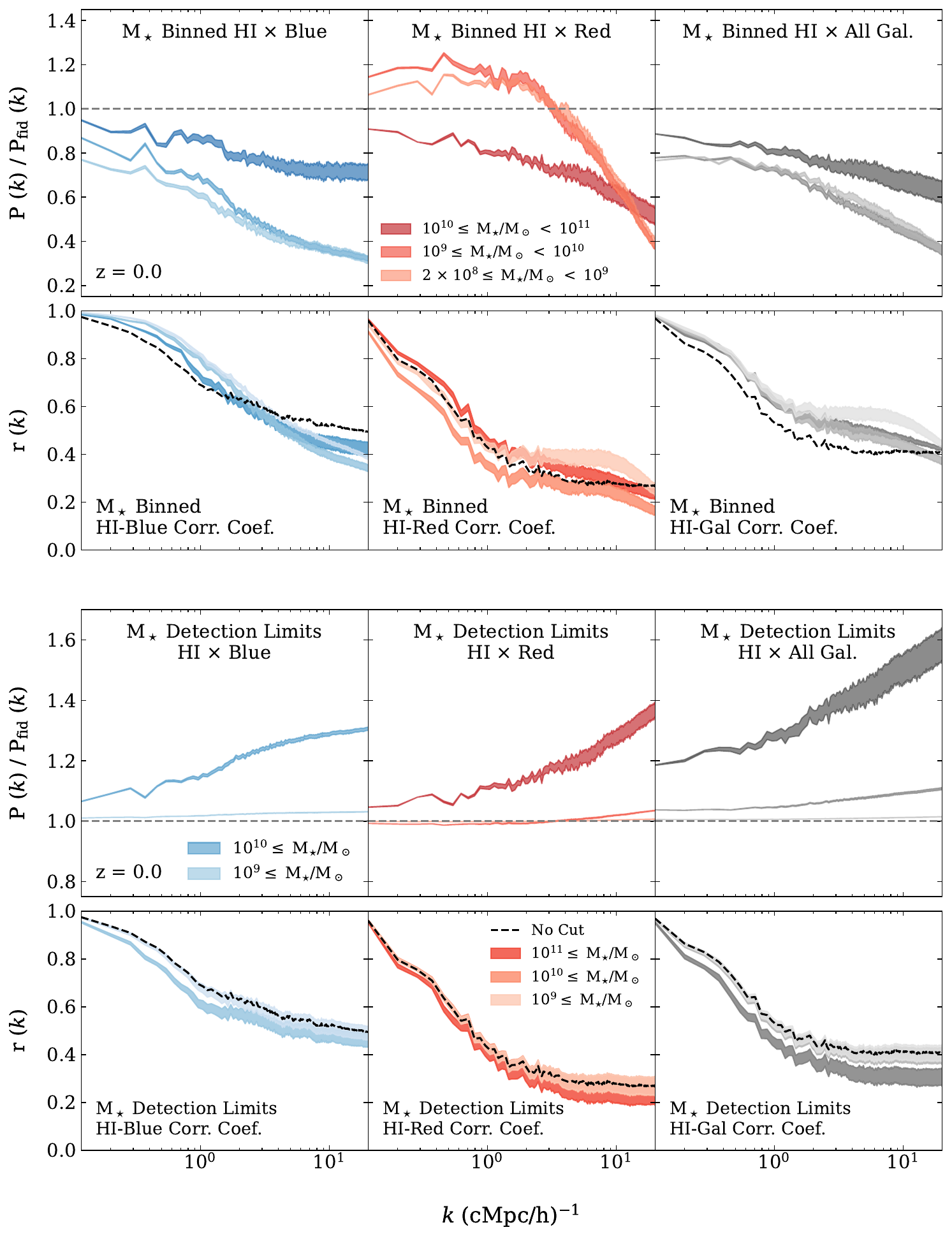}
    \caption{The clustering of stellar mass subsamples for \hib (left), \hir (centre), and \hia (right) in real space at $z = 0$. Contours encompass all \hi models, with darker colours representing subsamples with greater $M_\star$. The cross-power spectra for $M_\star$ bins (first row) and detection limits (third row) are shown as a ratio to the overall sample with no cuts in $M_\star$ for that galaxy colour. \hib and \hia both cluster more with increasing $M_\star$. \hir, however, clusters the most weakly in the largest mass bin (see text for discussion). The correlation coefficients for each galaxy colour and \hi (second row) evolve little with $M_\star$, demonstrating that differences in the cross-power spectra arise from changes in the clustering of the galaxy population rather than their spatial connection with \hi. In the bottom two rows, we display cross-power spectrum ratios and correlation coefficients for galaxy samples above certain $M_\star$ thresholds, which represent a rough proxy for detection limits. \hir and \hia converge to their fiducial cross-power with no (additional) mass cuts by $M_\star = 10^{10} M_\odot$ and \hib by $M_\star = 10^{9} M_\odot$. \hib converges at greater masses, as most galaxies in $M_\star \gtrsim 10^{10.5} M_\odot$ are red. We exclude the $M_\star \geq 10^{11} M_\odot$ threshold for \hib, as it is dominated by shot noise.}
    \label{fig:stmassbinth}
\end{figure*}

\begin{figure*}
    \centering
    \includegraphics[width=\linewidth, trim={0, 0.25cm, 0, 0}, clip]{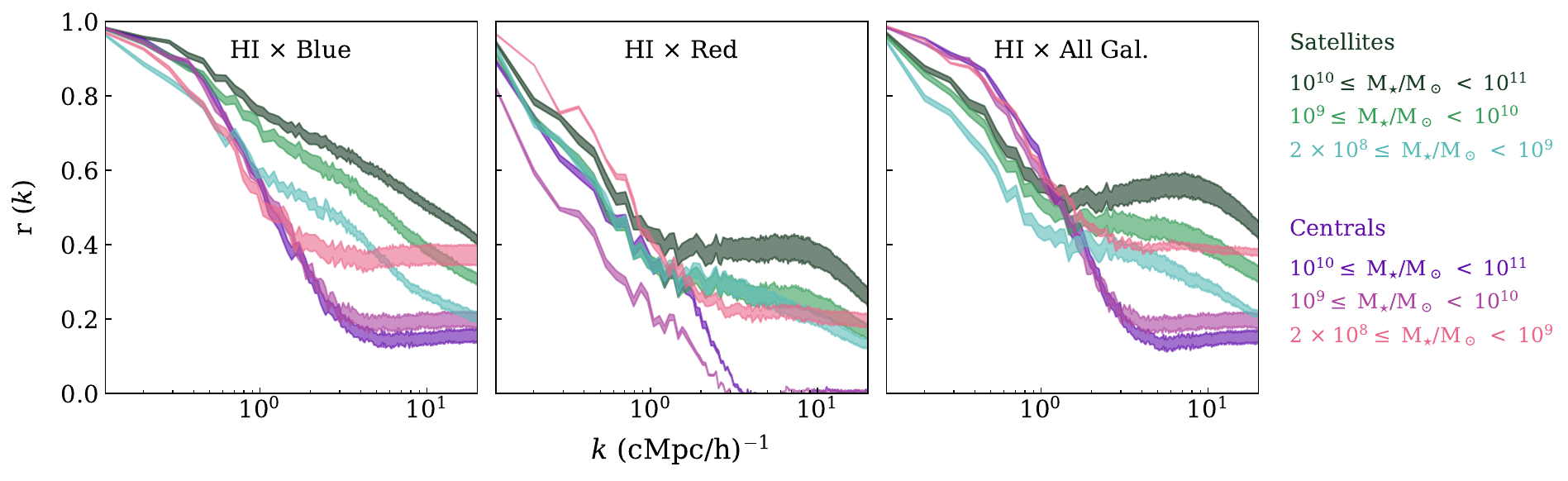}
    \caption{Correlation coefficients between \hi and three galaxy colour populations, blue (left), red (middle), and all galaxies (right) at $z = 0$. Each galaxy colour population is separated into stellar mass bins of width $\approx 1$ dex, with lighter colours representing less massive bins and darker more massive. Each stellar mass bin is further separated into centrals (purple) and satellites (green). \hi correlates more strongly with satellites for increasing $M_\star$ across galaxy colour, but this trend reverses for centrals. By definition, there is only one central galaxy per halo, thus the correlation coefficients for centrals are constant on small scales.}
    \label{fig:censat}
\end{figure*}


\subsection{Dependence on stellar mass in centrals and satellites} \label{subsection:stbinth}

It is well-established that more luminous or massive galaxies cluster more strongly \citep{ZehaviLumColCluster2005, ZehaviColGalClus2011, Beutler2013, Guo2013SDSS, GuoSDSS2014DR10, Skibba2014PRIMUSClustering, Zhai2023}. In this section, we examine how this relationship manifests in cross-correlations with \hi and test whether the tendency for observations to detect bright objects skews the measured cross-power spectra. We compare the clustering of galaxies whose stellar masses fall within three coarse $M_\star$ bins, as well as those whose stellar mass exceeds three lower detection limits with separations of $\approx 1$ dex. The $M_\star$ relationships of the cross-power spectra evolve little with redshift; for brevity, we examine only $z = 0$ results and provide similar analyses at other redshifts in the online figures.

In the top row of Fig.~\ref{fig:stmassbinth}, we separate blue, red, and all galaxies into coarse $M_\star$ bins and compute their cross-power spectra and correlation coefficients with \hi. The correlation coefficients distinguish changes in the spatial connection of the galaxy subsample to \hi from changes to the intrinsic clustering of the member distributions of the cross-power spectra (Section~\ref{subsection:HIcolor}). Both \hib (left column) and \hia (right) increase with $M_\star$, showing that more massive galaxies in the blue and whole galaxy samples cluster more strongly. \hir (middle) on the other hand does not exhibit a clear trend with $M_\star$; we will elaborate on this behaviour later in the section.

The correlation coefficients (second row) for all galaxy colours vary negligibly with $M_\star$. This lack of evolution implies that within each $M_\star$ bin, the correlation between blue or red galaxies and \hi does not significantly differ from that of other galaxies of the same colour but with different masses. Consequently, the $M_\star$ evolution of the cross-power spectra in the top row of Fig.~\ref{fig:stmassbinth} reflects the inherently stronger clustering of massive galaxies.

In addition to $M_\star$ bins, we also investigate cross-power spectra and correlation coefficients for lower $M_\star$ thresholds in Fig.~\ref{fig:stmassbinth}, which we consider a rough analogue for detection limits in observations. The cross-power spectra for \hir and \hia converge to the fiducial cross-power spectrum for the whole sample at $M_\star = 10^{10} M_\odot$, and \hib does the same at $M_\star = 10^{9} M_\odot$. \hi-galaxy cross-power spectra measured in observations with galaxy surveys complete up to those masses should be therefore insensitive to detection limits. The correlation coefficients for each $M_\star$ detection limit also converge to the fiducial sample by the same masses as the cross-power spectra for each colour.

The clustering of blue and all-galaxies in Fig.~\ref{fig:stmassbinth} increases with $M_\star$. Red galaxy clustering, however, only increases slightly from $2 \times 10^8 \leq M_\star / M_\odot < 10^9$ to $10^9 \leq M_\star / M_\odot < 10^{10}$, before decreasing drastically in the largest $M_\star$ bin. This trend aligns with observations and previous simulation work, where the red galaxy auto power spectrum decreases in the range $10^{8.5} \lesssim M_\star / M_\odot \lesssim 10^{10}$ before increasing in $10^{10} \lesssim M_\star / M_\odot \lesssim 10^{11.5}$ \citep{ZehaviLumColCluster2005, Guo2011, Henriques2017, SpringelTNG2018}. This unique trend is tied to the composition of each $M_\star$ bin with respect to centrals and satellites. The greatest $M_\star$ bin predominantly consists of centrals, whereas the two smallest mass bins are composed almost entirely of satellites. Central galaxies, by definition, must inhabit different haloes whereas satellites have no such restriction. Furthermore, satellites tend to inhabit massive halos which are strongly clustered. Consequently, red galaxies in the smaller $M_\star$ bins cluster more strongly than red galaxies in the largest $M_\star$ bins.

To investigate the effect of central/satellite demographics on the $M_\star$ evolution, we analyse the \hi correlation coefficients of blue, red, and all-galaxies split into centrals and satellites in Fig.~\ref{fig:censat}. \hia (right panel) correlation coefficients for centrals inherit their shape from \hib (left), as most centrals are blue. Similarly, most satellites are red, such that satellite correlation coefficients for \hia echo \hir. However, blue satellites boost the corresponding \hia coefficients by contributing some correlations with \hi. Central correlation coefficients (purple contours) approach constant values in the one-halo regime ($k \sim 0.3$ \hmpc) because only one central can occupy each halo.

For all galaxy populations, satellites correlate more with \hi with increasing $M_\star$ whereas centrals correlate less. We can interpret these correlations by understanding the influence of environment on a galaxy's gas abundance. Various external processes increase in strength and frequency with environmental density \citep{Brown2021VERTICO, Zabel2022VERTICO, Vicente2022VERTICO, Watts2023VERTICO}, such as starvation \citep{LarsonStarvation1980, vandevoort2017environmental}, ram pressure stripping \citep{GunnRPStrip1972, AbadiRPStrip1999}, and gas heating via satellite-satellite interactions \citep{MooreHarassment1996, MooreHarassment1998}, among others. Larger central galaxies tend to occupy denser haloes that are more effective at removing gas, and thus central galaxies correlate less with \hi with increasing $M_\star$ in the one-halo regime. Massive red centrals in particular (centre panel of Fig.~\ref{fig:censat}) are nearly completely uncorrelated with \hi out to $k \approx 3$ \hmpc, or $R = 2\pi/k \approx 2.1$ \mpch, demonstrating the efficacy of external quenching mechanisms near the centres of massive haloes \citep{GomezSFRvsEnv2003, BaloghColorLumEnv2004, Wolfquench2009, WooQuenchvsR2013}.  

\begin{figure}
    \centering
    \includegraphics[width=\linewidth]{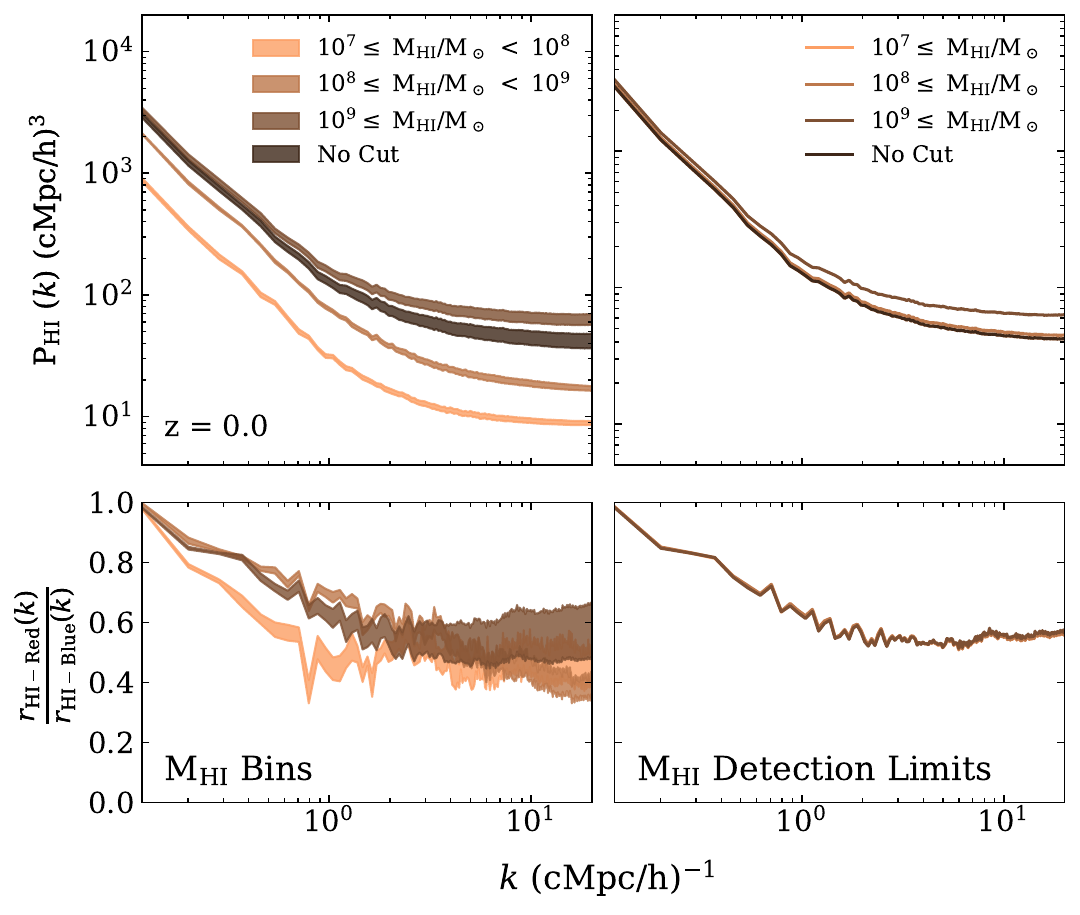}
    \caption{\hi auto power spectra of galaxies within a particular \mhi subsample at $z = 0$. For \mhi detection limits, we display only the centre of each contour encompassing the \hi models to avoid crowding the figure. Breaking up the galaxy population into \mhi bins (top left), we find that the \hi auto power grows with increasing \mhi (darker shades). The proximity of the fiducial \hi auto power spectrum (black) to the largest \mhi bin implies that high \mhi galaxies dominate \hi clustering. The similar auto power spectra for different \mhi thresholds (top right) demonstrate that the \hi auto power is insensitive to detection limits. We examine the influence of the colour make-up of each \mhi subsample by providing the correlation coefficient ratio between an \mhi bin/threshold and all red galaxies over blue galaxies. A smaller $r$ ratio indicates that the galaxies within particular \mhi bin/threshold correlate more with blue galaxies than other \mhi bins/thresholds. The similar $r$ ratios across \mhi subsamples indicate that these trends do not arise from evolving galaxy colour demographics but instead from the inherent clustering strength of galaxies with greater \mhi.}
    \label{fig:HIbinth}
\end{figure}

In contrast to centrals, satellite galaxies correlate \textit{more} with \hi with increasing $M_\star$. The reason for this trend is relatively straightforward for blue satellites. Massive blue satellites tend to inhabit larger \hi-rich haloes, resulting in stronger \hi correlations. On the other hand, the largest red satellites are typically hosted by haloes that are increasingly \hi-deficient at larger masses, as indicated by their low $\mhi / M_h$ \citep[see figure 4 in][]{V-NIngred21cm2018}. This characteristic should lead to a decrease in the correlation between red satellites and \hi at higher $M_\star$, which we do not find in Fig.~\ref{fig:censat}. 
We speculate that this discrepancy may arise from two effects: satellite-satellite correlations and resistance to environmental effects. Satellites in massive haloes are, relative to the rest of the halo, \hi-rich \citep[see figure 7 in][]{V-NIngred21cm2018}. Larger haloes contain a greater number of satellites, which may mitigate their \hi deficiency. The second effect, environment resistance, may also contribute as massive satellites possess deeper potential wells that can better withstand environmental stripping mechanisms \citep{Marasco2016EAGLEHI, StevensTNGHI2019, DonariAGNTNG2021}. 

\subsection{Dependence on galaxy H\textsc{i} mass} \label{subsection:hibinth}

In the previous section, we studied how the \hi-galaxy cross-power spectra evolve with $M_\star$. Here, we present a similar analysis with \mhi. Observations have reached different conclusions about the behaviour of galaxy clustering with \mhi. For example, \citet{BasilakosHIPASSpk2007} and \citet{Guo2017HImassClustering} claim that the \hi auto power spectrum increases as a function of \mhi, while \citet{MeyerHIPASSpk2007} and \citet{BigPapa2013} find no conclusive evidence of such a trend. We investigate this relationship in IllustrisTNG by separating galaxies at $z = 0$ into three coarse \mhi bins and thresholds separated by 1 dex, and measure the clustering and correlation strength with blue and red galaxies for each \mhi subsample in Fig.~\ref{fig:HIbinth}.

We find that galaxy clustering increases as a function of \mhi in the top left panel of Fig.~\ref{fig:HIbinth}, supporting the conclusions of \citet{BasilakosHIPASSpk2007} and \citet{Guo2017HImassClustering}. In the bottom row, we show the ratio of the \hi-red and \hi-blue correlation coefficients, $r\srm{HI-Red}/r\srm{HI-Blue}$, which describes how each \mhi subsample correlates with blue and red galaxies. \hi is expected to correlate more with blue galaxies than red galaxies; the ratio $r\srm{HI-Red}/r\srm{HI-Blue}$ indicates how much more a particular \mhi subsample correlates with blue galaxies over red galaxies. If lower \mhi bins are preferentially occupied by smaller and \hi-rich blue galaxies rather than larger but \hi-poor red galaxies, then the lower \mhi bins will exhibit a smaller $r\srm{HI-Red}/r\srm{HI-Blue}$ ratio. However, the correlation coefficient ratios evolve little between each \mhi bin, at most differing by a factor of $\sim 1.3$ on large scales. The small differences in $r\srm{HI-Red}/r\srm{HI-Blue}$ illustrate that blue and red galaxies have approximately equal influence over the clustering in each bin. This lack of evolution in our \mhi range implies that the \hi auto power spectrum increasing with \mhi arises from the tendency for galaxies to inhabit more massive haloes with increasing \mhi, rather than changes in the colour make-up within a particular \mhi bin. 

Although we conclude that galaxy clustering increases with \mhi, we emphasize that this does not necessarily contradict the findings of \citet{MeyerHIPASSpk2007} or \citet{BigPapa2013}. Both works conducted their analyses on galaxy samples in a higher \mhi regime, where we find some evidence that the trend of increasing \hi clustering may no longer hold (online figures). However, our box size is not sufficient to test the behaviour of \hi clustering at higher \mhi regimes.

We also test how different instrument detection limits will affect the measured \hi auto power spectrum, using \mhi cuts as a rough proxy. We present in the right panels of Fig.~\ref{fig:HIbinth} the clustering for subpopulations with cuts at $\mhi \geq 10^7 M_\odot$, $\mhi \geq 10^8 M_\odot$, and $\mhi \geq 10^9 M_\odot$. The \mhi-limited auto power spectra contains negligible changes, indicating that the \hi auto power spectrum is not sensitive to detection limits. The $r$ ratio for each threshold are identical (bottom right panel of Fig.~\ref{fig:HIbinth}). We conclude from this that \hi detection thresholds below $\mhi = 10^9 M_\odot$ do not cross-contaminate observations, in the sense that an instrument that detects galaxies down to that threshold will not be preferentially measuring \hi from \hi-rich blue galaxies or \hi-poor red galaxies.

\section{Discussion: What causes the redshift evolution of the power spectra?}
\label{section:discussion}

\begin{figure*}
    \centering
    \includegraphics[width = .75\linewidth]{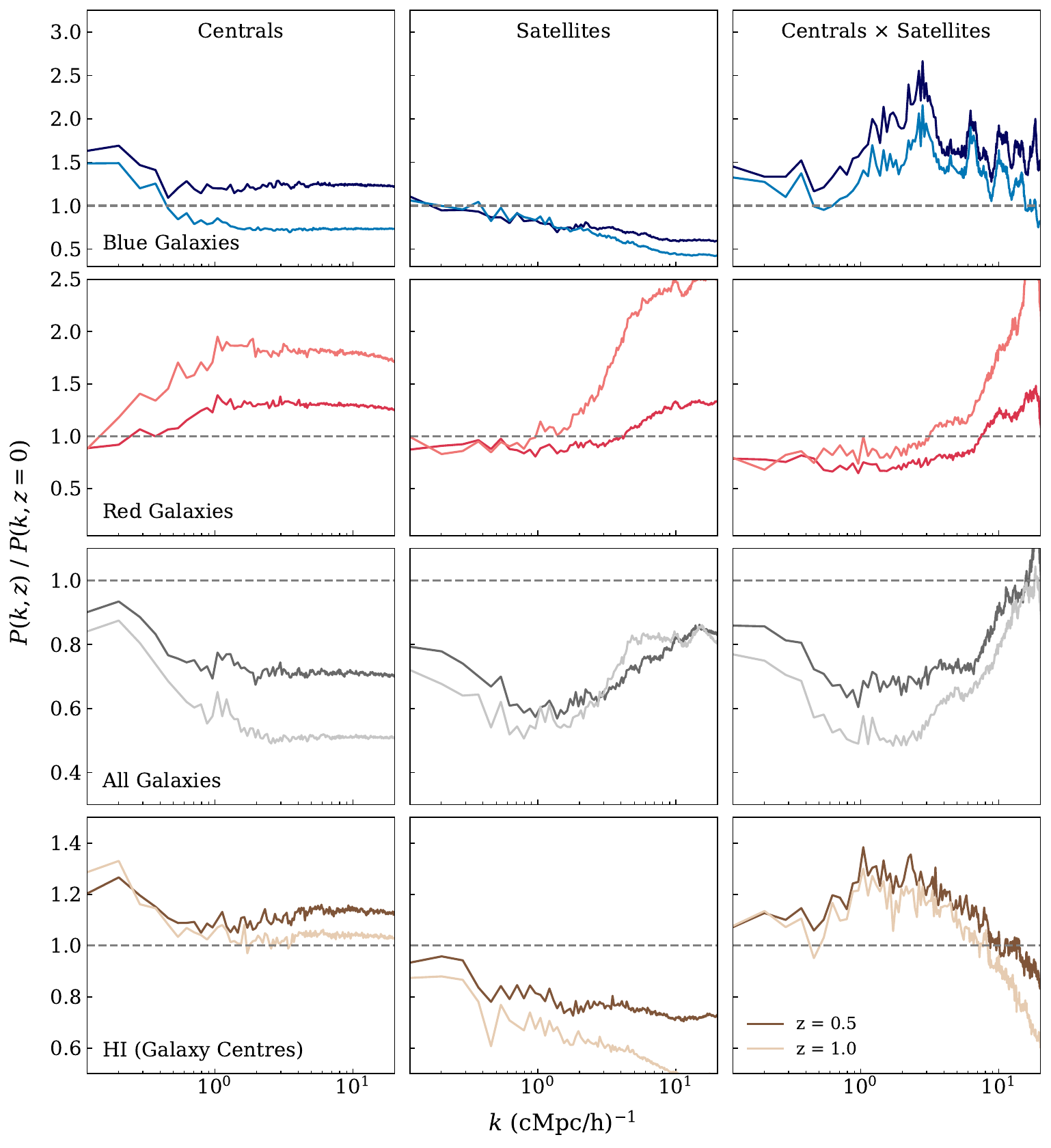}
    \caption{Redshift evolution of the blue (first row), red (second), all-galaxy (third), and \hi (last) auto power spectra split into contributions from centrals (\cen, left), satellites (\sat, centre), and their cross-correlation (\censat, right) according to equation~\ref{eq:pk_censat}. Power spectra at $z = 1$ (lighter colours) and $z = 0.5$ (darker) are normalised by the corresponding $z = 0$ power spectra, such that power spectra monotonically increasing from $z=1$ to $z=0$ approach unity from below and those decreasing from above. We employ the \textit{Galaxy Centres} \hi models to more clearly distinguish centrals and satellites in the \hi distribution, and display only the mean of the resulting power spectra for visibility. The blue galaxies and \hi power spectra lose a substantial contribution from the \censat term between $z = 0.5$ and $z = 0$, with a smaller loss from \cen. Red galaxies, on the other hand, lose considerable power on small scales from all terms, especially from \sat. These trends provide deeper insight into the processes responsible for the unexpected redshift evolutions found in Section~\ref{section:results} (see text).}
    \label{fig:censat_auto}
\end{figure*}

We have characterized the clustering of \hi as a function of colour, redshift, $M_\star$, and \mhi, often alluding to the influence of baryonic processes on these relationships. In particular, we attributed much of the inverse redshift evolution of the power spectra from Section~\ref{sub:zevo} to baryonic processes, but did not determine the precise mechanisms responsible and scales at which they take place. Previous studies of the redshift evolution of red galaxy clustering have proposed mergers, disruption, galaxy quenching and some combination thereof as possible mechanisms that suppress structure growth in galaxies \citep{zehavi2005LRGzevo, White2007, Guo2013SDSS}. In our analysis, we have introduced colour transitions and gas loss as alternative suppression mechanisms. We attempt to disentangle these processes in the redshift evolution of the power spectra in Section~\ref{sub:zevo} by considering the clustering of smaller galactic subpopulations. We first decompose the power spectra into contributions from centrals and satellites. Next, we describe how colour transitions and gas loss manifest in each component. We finish by briefly describing whether the redshift evolution of these terms can be explained with other processes, such as mergers and galaxy interactions.

We separate the blue, red, all-galaxy, and \hi auto powers into contributions from centrals and satellites to provide insight into the processes that govern their time evolution, shown in Fig.~\ref{fig:censat_auto}. We use \textit{Galaxy Centres} for the \hi distribution in order to more clearly distinguish between centrals and satellites, but this should have little effect on our conclusions here (Section~\ref{subsection:HIcluster}). \citet{OkumuraCSHOD2015} provide the following framework for separating the galaxy auto power spectra into central and satellite components,
\begin{equation} \label{eq:pk_censat}
    P(k) = (1 - f_s)^2 \cen (k) + 2f_s(1 - f_s)\censat (k)+ f_s^2 \sat (k) \,,
\end{equation}
where the contributions from centrals (c) and satellites (s) are denoted in subscripts and $f_s$ represents the satellite fraction. 
Each term from equation~\ref{eq:pk_censat} contributes differently at various spatial scales, depending on the shape of the term itself and the corresponding $f_s$ coefficients \citep[see figure 1 in ][]{OkumuraCSHOD2015}. By definition, there is only one central per halo, so \cen is dominated by shot noise in the one-halo regime. The red galaxy auto power spectrum, however, is dominated by \sat on all scales because of the high satellite fraction for red galaxies ($f_s \sim 0.8$ in TNG100). In contrast, the small satellite fraction for blue galaxies ($f_s \sim 0.3$) significantly down-weights \sat. We now study the evolution of the terms from equation~\ref{eq:pk_censat} in Fig.~\ref{fig:censat_auto}.

Every term from the all-galaxies power spectra increases with time (third row), retaining the structure growth seen in the full power spectrum from Section~\ref{sub:zevo}, whereas the evolution in other tracers is significantly more nuanced. Each component of the blue galaxies (top row) and \hi (bottom row) increases slightly between $z = 1$ and $z = 0.5$ in Fig.~\ref{fig:censat_auto}, before both of the central-dependent terms (\cen and \censat) decrease between $z = 0.5$ and $z = 0$. Red galaxies (second row), on the other hand, cluster less with time in every term on small scales. This effect is particularly strong in the \sat term, where the clustering of satellites falls by a factor of 2.5 from $z = 1$ to $z = 0$.

We now analyse how colour transitions affect the previously described evolutions from Fig.~\ref{fig:censat_auto}. As described in Section~\ref{sub:zevo}, colour transitions suppress clustering by adding relatively weakly clustered galaxies to the red population and removing strongly clustered galaxies from blue. The impact of colour transitions is most visible in the suppression of the clustering of blue centrals (top right panel of Fig.~\ref{fig:censat_auto}), suggesting transitioning blue centrals are primarily responsible for the decreasing power spectrum between $z = 0.5$ and $z = 0$. The suppression in \cen (left) is strongest at largest scales, where transitioning blue centrals dominate because they tend to occupy the most massive blue-hosting haloes. However, in the case of \censat (right), the massive haloes hosting transitioning centrals contribute substantially more to smaller scales because they possess the highest satellite fractions amongst blue-hosting haloes. Consequently, when all the central-satellite pairs from that halo are removed after their central transitions, \censat decreases significantly at the boundary between the one-halo and two-halo regimes ($k \sim 2$ \hmpc). Previous studies of galaxy quenching in IllustrisTNG indicate that AGN feedback is largely responsible for the colour transitions in these centrals, particularly when their mass reaches the sharp transition at $M_\star \geq 10^{10.5} M_\odot$ \citep{ZingerAGNTNG2020,DonariAGNTNG2021}. The last term, \sat (middle), still increases with time, indicating that colour transitions are less effective at removing power from blue satellites.

In the case of the red population, transitioning galaxies influence their clustering on small scales rather than large scales. As depicted in the middle row of Fig.~\ref{fig:censat_auto}, \cen (left) is suppressed moderately at the smallest scales of the two-halo regime ($k \sim 1$ \hmpc), while \sat and \censat decrease drastically almost entirely within the one-halo regime. Transitioning galaxies inhabit the smallest and youngest red-hosting haloes, which contribute most to small-scale red galaxy clustering, as large scales are dominated by the largest haloes. This is particularly true for the clustering of satellites, as new red-hosting haloes have smaller satellite fractions and satellite quenched fractions \citep{DonnariQuenchObs2021}.

The other baryonic process we examine is gas loss, which only directly impacts the clustering of \hi but not of galaxies. The evolution of \hi auto power spectrum components (bottom row) echo the corresponding blue terms (top row), although with a few additional subtleties. When a galaxy transitions from blue to red, it is completely removed from the blue distribution. The associated gas loss, however, simply down-weights previous contributions of that galaxy to the \hi auto power spectrum. This mitigates the inverse evolution of the \cen (left) and \censat (right) terms at $z = 0.5$ to $z = 0$ as compared to blue galaxies. Similarly to the effect of colour transitions on blue galaxies, we attribute the abating \hi power spectrum primarily to the rapid gas loss in and around centrals. The allocation of \hi between centrals and satellites in massive haloes further supports this conclusion. \citetalias{V-NIngred21cm2018} find that satellites contain most of the \hi in $M_h \gtrsim 10^{13} M_\odot$ haloes at $z \leq 1$ (see their figure 7). The \hi profiles of these haloes provide further evidence for gas loss in and around centrals. The profiles of $M_h \sim 10^{13} M_\odot$ haloes contain $\sim 10$ kpc / \textit{h} ``holes'' around their centres which steadily deepen and broaden with time, a characteristic also found in massive galaxies from other simulations \citep{Bahe2016EAGLEHI, Stevens23}. At $z = 0$, $M_h \sim 10^{12} M_\odot$ haloes also develop similar features in their profiles (see figure 5 from \citetalias{V-NIngred21cm2018}), showing an increased prevalence of this characteristic at later times even at lower-mass haloes. This phenomenon supports our conclusion that gas loss near the centres of previously gas-rich haloes suppresses the growth of the \hi auto power spectrum.

Colour transitions and gas loss do not comprise an exhaustive list of mechanisms that can suppress power spectra. Mergers and disruptions, for example, can also impede the rate of clustering growth in the one-halo regime by reducing the number density of satellites \citep{Bell2004, Skelton2009, Guo2013SDSS}. However, mergers are unlikely to elicit the decrease in blue galaxy clustering, as blue galaxies merge at a relatively slow rate \citep{linmerger2008, Darg2010}. Red galaxies appear to follow this description in the \sat and \censat terms, which decrease significantly at small scales. However, the all-galaxy \sat term \textit{increases} with time at $k \sim 5$ \hmpc by a factor of $\sim 1.5$, despite red galaxies decreasing by a factor of $\sim 2$ at the same scale. Mergers should affect both the red and all-galaxy \sat terms \citep{watson_mergers_corrfunc2011}, particularly at $z = 0$ where red galaxies comprise $\sim70\%$ of the mass in satellites. However, we note that our reasoning throughout this section neglects the impact of galaxies crossing our imposed resolution limit and centrals becoming satellites between redshifts. Properly disentangling the contributions of various baryonic processes requires sophisticated modelling \citep{Guo2013SDSS}, which we leave for future work.

In summary, we find evidence that previously neglected baryonic processes, namely colour transitions and gas loss, significantly influence the clustering of blue and red galaxies and \hi at $z \leq 1$ such that the power spectra for these populations decrease with time. We identify the signatures of colour transitions and gas loss in the evolution of the terms from equation~\ref{eq:pk_censat} in Fig.~\ref{fig:censat_auto}. These processes in (previously) gas-rich and star-forming haloes appear particularly effective at suppressing the clustering of each population.

\section{Conclusions}
\label{section:conclusion}

We have presented the first systematic investigation of the cross-power spectra of \hi and galaxies split into colour subpopulations in IllustrisTNG and studied how their clustering changes with time and various galaxy properties. We find the following:

\begin{enumerate}
    \item The clustering of simulated \hi distributions exhibits only a weak dependence on the model for the transition between atomic and molecular hydrogen.
    
    \item The \hi-red galaxy cross-power spectrum (\hir) is greater than \hi-blue (\hib) at large scales due to red galaxies' inherent clustering strength and larger bias with respect to matter. However, processes such as AGN feedback and ram-pressure stripping suppress \hi abundance in the massive haloes that red galaxies typically occupy, weakening \hir on small scales and creating an intersection between \hir and \hib at $\approx 3$ Mpc at $z = 0$. 
    
    \item In redshift space, the suppression of power due to the fingers-of-God effect manifests more strongly in red galaxies than blue galaxies, introducing a secondary colour-dependency and pushing the intersection of \hir and \hib to $\approx 10$ Mpc at $z = 0$. 
    
    \item The \hi, red, and blue galaxy auto power spectra and their cross-power spectra \textit{decrease} with cosmic time, contrary to the clustering of matter and the galaxy population as a whole. Colour transitions in galaxies and \hi consumption contribute to this inverse evolution. These baryonic processes may need to be taken into account in models of the large-scale distribution of these populations at $z < 1$.

    \item \hib increases as a function of $M_\star$. \hir also reflects this trend, until $M_\star \approx 10^{10} M_\odot$ where the clustering is the weakest amongst the stellar mass bins examined. Red galaxies below this threshold are typically satellites that occupy massive haloes more frequently than the red centrals in larger $M_\star$ bins. 
    
    \item Satellites correlate more strongly with \hi with increasing $M_\star$, whereas centrals correlate less strongly. These trends hold regardless of galaxy colour.

    \item The \hi clustering increases as a function of \mhi, and galaxies with $\mhi \geq 10^8 M_\odot$ dominate the total \hi auto power spectrum. We show that \hi auto power spectra should be unbiased as long as the survey captures galaxies with $\mhi \geq 10^8 M_\odot$.   
\end{enumerate}

These conclusions are important for future 21cm surveys, where detections occur in the first phases as cross-correlations. These results contribute to the theoretical formalism needed to extract cosmological constraints from upcoming 21cm surveys and better understand the \hi-galaxy-halo connection. We have established that the baryonic processes associated with quenching can have large-scale imprints on the clustering of \hi, blue, and red galaxies. Models of the bias with respect to matter of these populations rely on simplistic assumptions about their growth and scale-dependence, which may skew the interpretation of the data from 21cm surveys, as we will explore in future work (Osinga et al., in prep.).

One caveat to our results is that they were derived from a single suite of simulations, IllustrisTNG. Repeating our analysis with other simulations could illuminate whether these relationships are sensitive to the simulation's model for galaxy formation and evolution. For example, it is unclear whether or not the cosmic abundance of \hi ($\Omega\srm{HI}$) in IllustrisTNG is consistent with observations in the redshifts studied here \citep{V-NIngred21cm2018, DiemerHI2019}. Furthermore, mock-observing IllustrisTNG would allow for more faithful comparisons with observations and further our understanding of how observational effects manifest in the clustering relationships examined here. For example, \citet{DonnariQuenchObs2021} showed that quenched fractions are quite sensitive to observational effects.

\section*{Acknowledgements}
\label{section:acknowledgements}

We would like to thank Alberto Bolatto and Massimo Ricotti for their guidance during the project, and Adam Stevens for the useful discussions. Research was performed in part using the compute resources and assistance of the UW-Madison Centre For High Throughput Computing (CHTC) in the Department of Computer Sciences. The CHTC is supported by UW-Madison, the Advanced Computing Initiative, the Wisconsin Alumni Research Foundation, the Wisconsin Institutes for Discovery, and the National Science Foundation. We also acknowledge the University of Maryland supercomputing resources (http://hpcc.umd.edu) made available for conducting the research reported in this paper. Other software used in this work include: \textsc{numpy} \citep{numpy2020}, \textsc{matplotlib} \citep{mpl2007}, \textsc{seaborn} \citep{seaborn2021}, \textsc{h5py} \citep{h5py2021}, \textsc{colossus} \citep{colossus18} and \textsc{pyfftw} \citep{gomersall2021pyfftw}.

\section*{Data Availability} \label{section:acknowledgements}

Data from the figures in this publication are available upon reasonable request to the authors. The IllustrisTNG data is available at \url{www.tng-project.org/data}.


\bibliographystyle{mnras}
\bibliography{refs} 



\appendix

\section{Sensitivity to colour cut}
\label{appendix:color_cut}

\begin{figure}
    \centering
    \includegraphics[width = .65\linewidth]{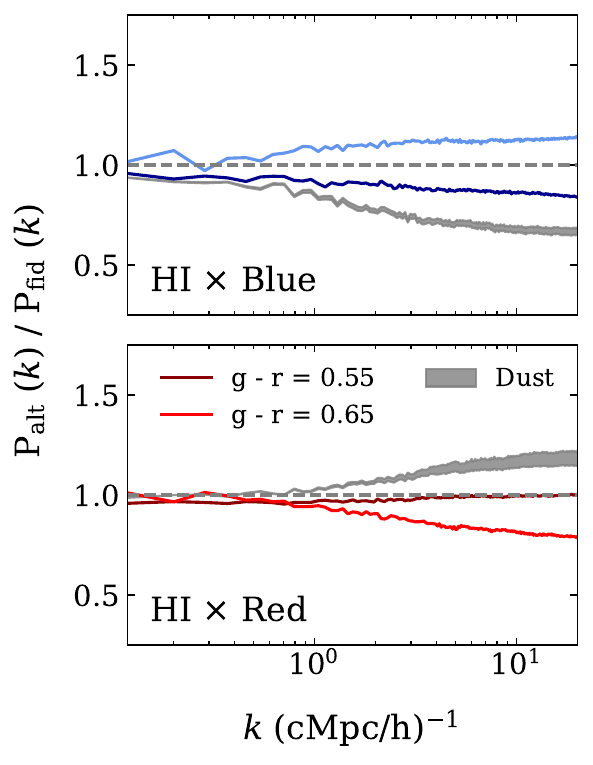}
    \caption{The sensitivity of \hib (top) and \hir (bottom) to the colour cut and dust (gray) at $z = 0$. The cross-power spectra calculated with colour cuts at two extremes \gr = 0.55 (dark) and \gr = 0.65 (light) are shown as a ratio over the fiducial \gr = 0.60 cut. We apply \gr = 0.60 for the dust prescription from \citet{NelsonTNG2018}. We plot the centre of the contours encompassing \hi models from the alternate \gr cuts for visibility, but they have similar widths to the dust contour. Particularly in the two-halo regime, power spectra are insensitive to the colour threshold. On smaller scales, the values of the power spectra are less important since we focus on trends to gain insight into galaxy evolution.}
    \label{fig:cuts}
\end{figure}

Throughout the paper, we have used a single cut in \gr to separate blue and red galaxies, although this cut changes with time. Here, we analyse the sensitivity of the cross-power spectra to alternate colour cuts and the inclusion of dust \citep{NelsonTNG2018}. We particularly focus on the effect on large scales ($k \lesssim 1$ \hmpc), since these are the scales used to constrain cosmological quantities and are what we compare to observations. On small scales, we have focused on trends rather than exact quantities and thus are less concerned with sensitivity to systematics.

The alternative colour definitions for \hir and \hib at $z = 0$ are displayed as a ratio over the fiducial colour cut in Fig.~\ref{fig:cuts}. \hib differs by $\lesssim 5\%$ on large scales ($k \lesssim 1$ \hmpc) and \hir changes negligibly. We intend for the alternative colour cuts to represent extremes, considering that they move away from the fiducial cut by 0.05 or $\sim 7.5\%$ of the entire range of \gr values. Consequently, we consider the $\lesssim 5\%$ change in \hib reassuring rather than concerning.


Dust attenuation reddens blue galaxies more than red ones, essentially squeezing the galaxy distribution along the \gr axis. This moves the minimum between the blue and red modes in the \gr distribution from \gr  = 0.6 to \gr $\approx 0.64$. We still use the fiducial \gr = 0.6 and consider the $\lesssim 10\%$ change in \hib on large scales as an upper limit. On small scales, dust effects are stronger, with \hib falling below the fiducial definition by a factor of $\sim 0.6$ and \hir increasing by a factor of $\sim 1.2$. However, on these scales we focus on trends to yield insight into galaxy evolution which are unaffected by the inclusion of dust. As long as the colour cut reasonably splits the colour modes, the clustering is comparable.

\section{Convergence Tests}
\label{appendix:convergence}

In this section, we ascertain the convergence of our results. We first test convergence of \hir and \hib with mass resolution, again focusing on large scales where the precise quantities of the power spectra are important (see Appendix \ref{appendix:color_cut}). Mass resolution will affect our results through three known avenues.

First, decreasing simulation resolution will increase the mass of our resolution limit of 200 particles ($2 \times 10^8 M_\odot$, $10^9 M_\odot$, and $9 \times 10^9 M_\odot$ for TNG100-1, TNG100-2, and TNG100-3, respectively). This removes low-$M_\star$ galaxies from the blue and red populations and galaxies with low $M_\star$ \textit{and} $M_{\mathrm{gas}}$ from the \textit{Galaxy Centres} and \textit{Particles in Galaxies} \hi distributions. We do not expect this first effect to substantially change our results. In Section \ref{subsection:stbinth}, we found that increasing $M_\star$ thresholds does not affect galaxy clustering until $10^{10} M_\odot$ for all galaxy colours. The \hi auto power spectra also do not depend on the increasing resolution cuts as the \hi auto power spectrum from \textit{All Particles}, which does not have any resolution cuts, matches the \textit{Galaxy Centres} and \textit{Particles in Galaxies} power spectra.

Second, the stochastic sampling of a galaxy's SFR can make low-mass galaxies that should be star-forming appear to be quiescent \citep{Trayford18, NelsonTNG2018}. The lower SFR for low-mass galaxies has several ramifications that are difficult to study in isolation, but \citet{NelsonTNG2018} found that the population of blue galaxies with stellar masses in the range $10^{9.5} M_\odot < M_\star < 10^{10} M_\odot$ is $\sim 0.67$ times smaller in TNG100-2 than in TNG100-1 (see their figure A3). This is particularly true when considered together with the third avenue for mass resolution dependence: uncalibrated model parameters. The models for IllustrisTNG are tuned to reproduce a set of observables in TNG100-1 \citep{PillepichTNGmodel2018}, and these parameters are left unchanged for coarser-resolution versions. This choice was made for easier convergence studies, but the results of subgrid models may be resolution-dependent, leading to different galaxy formation outcomes.

\begin{figure}
    \centering
    \includegraphics[width=.65\linewidth]{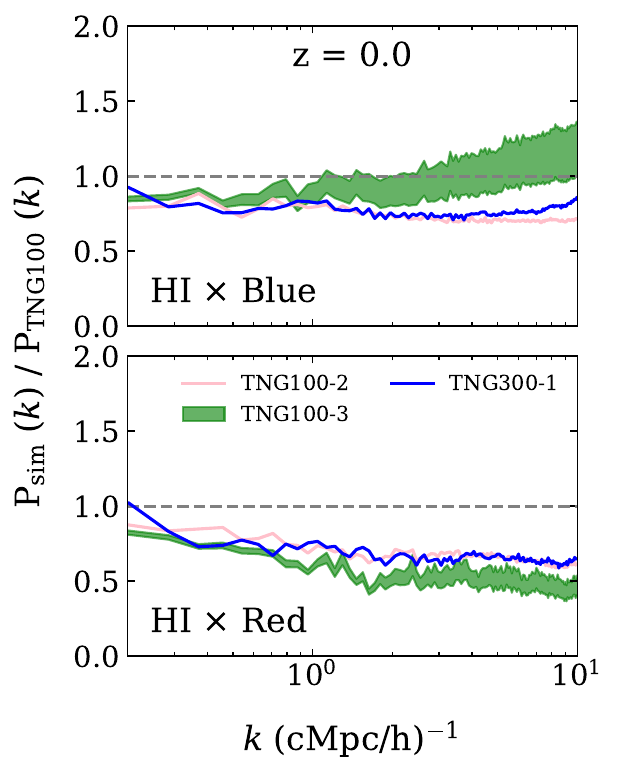}
    \caption{Convergence of \hib (top) and \hir (bottom) with mass resolution at $z = 0$, shown as a ratio over the fiducial TNG100 simulation resolution. The lower resolution versions of TNG100, TNG100-2 (pink) and TNG100-3 (green) are displayed in fainter colours. For visibility, we only plot the contour delineating \hi models for TNG100-3, but the widths are otherwise similar. In the overlapping $k$ range TNG100-2 and TNG300-1 agree for both \hib and \hir, which is reassuring since they have roughly the same mass resolution.}
    \label{fig:box}
\end{figure}

\begin{figure}
    \centering
    \includegraphics[width = .75\linewidth]{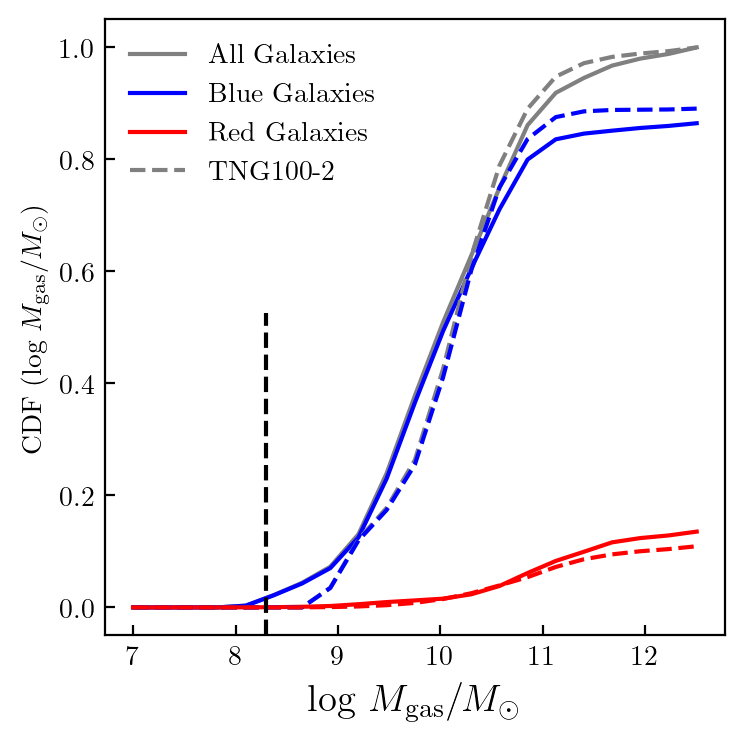}
    \caption{CDF of \hi in TNG100 (solid) and TNG100-2 (dashed) amongst $z = 0$ galaxies across different galaxy colours using $M_{\mathrm{gas}}$ bins with widths of $\sim 0.25$ dex. The vertical black dashed line shows the TNG100-1 gas mass cut. Galaxies below this resolution cut contain $< 3\%$ of the \hi from TNG100-1, suggesting that further resolution improvements will not have substantial impacts on our results.}
    \label{fig:HI_gas}
\end{figure}

We expect resolution-dependence in our results because of the last two effects. Many other works that use IllustrisTNG or other hydrodynamic simulations also find significant deviations and non-convergence in quantities dependent on SFR or gas content when comparing different mass resolutions \citep[e.g.,][]{Trayford18, NelsonTNG2018, DiemerHI2018,V-NIngred21cm2018, StevensTNGHI2019}. We also find that \hib and \hir only converge to within a factor of 2 in Fig. \ref{fig:box}. \hib is affected the least, remaining within a factor of $\sim 0.8$ of the TNG100-1 result across different scale regimes, which agrees with the mass resolution dependence found in the \hi auto power spectra from \citet{V-NIngred21cm2018}. \hir, however, can change by a factor of $\sim 2$ even on large scales, arising from disparate resolution-dependent behaviour between \hi and red galaxies. The auto power spectrum for red galaxies is nearly identical in TNG100-1 and TNG100-2, so the resolution-dependence of \hi clustering further severs the spatial connection between \hi and red galaxies for lower resolutions (see online figures for resolution dependence of auto power spectra).

Nonetheless, the resolution issues do not significantly affect our conclusions. First, TNG100-3 has nearly two orders of magnitude worse resolution than TNG100-1, and nearly one above TNG100-2. Given the complicated resolution-dependent behaviour described above, finding no resolution dependence or uniform resolution convergence would have been surprising. Second, TNG100-1 agrees with relevant observed quantities such as the \hi halo relation \citep{Obuljen2018} and blue and red galaxy clustering \citep{SpringelTNG2018} among others, implying that the insights derived from our results are relevant to observations. Third, we find some evidence indicating that further resolution increases may not result in substantial changes to our results. Fig. \ref{fig:HI_gas} shows the cumulative distribution functions (CDF) of the \hi from galaxies of different colours and in different $M_{\mathrm{gas}}$ bins. The slope of the TNG100-2 CDF stays relatively steep from $M_{\mathrm{gas}} \approx 10^{10.5} M_\odot$ until its resolution limit at $M_{\mathrm{gas}} \approx 10^{9} M_\odot$, which demonstrates that a cosmically significant amount of \hi occupies galaxies in $M_{\rm{gas}}$ bins around the TNG100-2 cutoff. This is confirmed by the TNG100-1 CDF, which has $\sim 10\%$ of its \hi is found in $M_{\mathrm{gas}} < 10^{9} M_\odot$ galaxies. The TNG100-1 CDF levels out between the TNG100-1 and TNG100-2 cutoffs, suggesting that galaxies in even smaller $M_{\rm{gas}}$ bins house an even less significant fraction of the cosmic \hi than the galaxies between the TNG100-1 and TNG100-2 cutoffs. Without simulations with higher resolutions, we cannot claim with certainty that the \hi occupation in TNG100-1 is converged, but we find Fig. \ref{fig:HI_gas} reassuring.


We also test how our results change with increasing grid resolution. We binned mass distributions in $800^3$, $1000^3$, and $1200^3$ grids and found no visible difference between them. This comparison is provided in the online figures. 

Since we arbitrarily picked a line of sight to project the velocities of matter on, we also projected our redshift-space power spectra along different axes. The cross-power spectra calculated from distributions displaced along different axes should not necessarily match, as matter will collapse faster along one axis than the others \citep{zeldovich1970}. Our results on large scales can vary by a factor of $\sim 15\%$ due to the axis chosen. This comparison is also provided in the online figures.


\bsp	
\label{lastpage}
\end{document}